\documentclass[aps,pre,reprint,superscriptaddress]{revtex4-2}

\usepackage{lipsum}
\usepackage{color}
\usepackage{amsmath}
\newcommand\ddfrac[2]{\frac{\displaystyle #1}{\displaystyle #2}}
\newcommand{\comment}[1]{}
\usepackage{listings}
\usepackage{float}
\usepackage[caption=false]{subfig}
\usepackage{graphicx} 
\usepackage{hyperref}
\usepackage{interval}
\usepackage{longtable}
\graphicspath{ {./figures/} }
\begin{document}



\title{Non-Gaussian distribution of displacements for Lennard-Jones particles in equilibrium}


\author{Aleksandra Pachalieva}
\email[]{apachalieva@lanl.gov}
\affiliation{Center for Nonlinear Studies, Los Alamos National Laboratory, Los Alamos, NM 87545, USA}
\affiliation{Department of Mechanical Engineering, Technical University of Munich, 85748 Garching, Germany}

\author{Alexander J. Wagner}
\email[]{alexander.wagner@ndsu.edu}
\affiliation{Department of Physics, North Dakota State University, Fargo, ND 58108, USA}

\date{\today}

\begin{abstract}
Most meso-scale simulation methods assume Gaussian distributions of velocity-like quantities. These quantities are not true velocities, however, but rather time-averaged velocities or displacements of particles. We show that there is a large range of coarse-graining scales where the assumption of a Gaussian distribution of these displacements fails, and a more complex distribution is required to adequately express these distribution functions of displacements.
\end{abstract}


\maketitle

\section{Introduction}

A key assumption of many meso-scale simulation methods is that the particle displacements follow a Gaussian distribution. This is motivated from the well known Gaussian distribution of velocities in equilibrium which is true for Brownian Dynamics \cite{ermak1978brownian}, Dissipative Particle Dynamics \cite{hoogerbrugge1992simulating}, Stochastic Rotation Dynamics \cite{ihle2001stochastic}, and the lattice Boltzmann method \cite{qian_lattice_1992,he1997theory}.

When re-deriving these methods from coarse-graining of Molecular Dynamics (MD) simulations, it becomes clear that these methods deal with finite time displacement distributions rather than velocity distributions. This has been explored for the lattice Boltzmann method by Parsa \textit{et al.} in a number of recent publications on the Molecular-Dynamics-Lattice-Gas (MDLG) method \cite{parsa_lattice_2017,parsa_validity_2019,parsa2020}. The MDLG method maps a coarse-graining of a Molecular Dynamics simulation onto an integer lattice gas. The MDLG analysis is a first principles approach to analyze Lattice Gas (LG) and Lattice Boltzmann Methods (LBM). It shows a great promise to develop a better understanding of fluctuating \cite{parsa2020}, thermal, multiphase and multicomponent lattice gas and lattice Boltzmann methods.

In previous MDLG studies by Parsa \textit{et al.} \cite{parsa_lattice_2017, parsa_validity_2019}, it was similarly assumed that the distribution function of the displacements is described by a Gaussian distribution function. The authors matched the second order moment to the theoretical mean squared displacement, which appeared to adequately predict the global equilibrium distribution function of a LBM. 

While trying to derive a universal LBM collision operator from the MDLG method, we examine a non-equilibrium system that undergoes simple shear flow. In deriving the hydrodynamic limit of a lattice Boltzmann method it is universally assumed that collisions keep the distribution functions close to local equilibrium, constrained by the conserved quantities, usually mass and momentum. We find that this is indeed the case. However, we observed that the distributions fail to approach local equilibrium, and we noticed that there are typically small errors (approx. 5\%) between the predicted and the measured equilibrium distribution functions. 

Since the only assumption that goes into the calculation of a local equilibrium in the MDLG approach is the distribution of displacements \cite{parsa_lattice_2017}, we had to conclude that the distribution of displacements must differ from a Gaussian distribution. This observation motivated the current study of the distribution of displacements in Molecular Dynamics simulations.

The question of physical displacements of particles has not received a lot of attention, but is also of general interest in Statistical Mechanics, as the short-term displacement is often modelled by a random walk. This has been discussed recently by Masoliver \textit{et al.} \cite{masoliver_three-dimensional_2017,masoliver_two-dimensional_2020} in the sense of the telegrapher's equation.

The paper is structured as follows: In Section\,\ref{sec:motivation}, we show the numerical evidence that the distribution of displacements indeed differs from a Gaussian distribution. This is followed by a detailed description of the simulation setup used to obtain the MD data given in Section\,\ref{sec:setup}. In Section\,\ref{sec:gaussian}, we show the mismatch between the MD data and the single Gaussian distribution of displacements. We propose two novel probability distribution functions which could be adjusted to match the second and fourth order moments of the measured data, respectively in Sections\,\ref{sec:ball_diff} and \ref{sec:poisson_gaussians}. Finally, some concluding remarks and future work are mentioned in Section\,\ref{sec:conclusions}.
\section{Motivation}
\label{sec:motivation}
In typical hydrodynamic systems, the locally conserved quantities are relaxed towards global equilibrium much faster than quantities that can be relaxed through collisions. For these systems the distribution of particle velocities will be close to a Maxwell-Boltzmann distribution corresponding to the local conserved quantities density, momentum, and temperature. This observation is at the core of many descriptions of non-equilibrium thermodynamics. For the Boltzmann equation it leads to an approximation which allows the two-particle collision term to be replaced by a simpler term of relaxing the velocity distribution towards the local Maxwellian distribution. This is known as the Bhatnagar-Gross-Krook (BGK) approximation \cite{bhatnagar_model_1954}. In the BGK formalism, the entire local relaxation depends on the details of small deviations from the local equilibrium distribution function.

In the MDLG 
context, we measure the distribution function of particle displacements from an underlying MD simulations of Lennard-Jones particles in equilibrium and thus, obtain an equilibrium distribution function for a specific simulation. For the particular application of measuring collisions, it is required to obtain precise measurements of the deviations from equilibrium. We noticed that the collision operator did not appear to relax towards the equilibrium distribution function predicted by Parsa \textit{et al.} \cite{parsa_lattice_2017}, but instead it relaxes to a distribution that deviates by a few percent. This deviation was not previously noticed but since now we were examining small deviations from equilibrium, these differences between the predicted and measured equilibrium distributions have the same order of magnitude as the non-equilibrium contributions to the distribution function. Since the only ingredient in the analytical prediction of the MDLG equilibrium distribution is the distribution of particle displacements \cite{parsa_lattice_2017}, we began to question the validity of the assumption that the distribution of the local displacements was truly given by a Maxwell-Boltzmann distribution, as expected.

\begin{figure}[htbp!]
     \centering
     \subfloat[]{\label{subfig-1:df_g}{\includegraphics[width=\columnwidth]{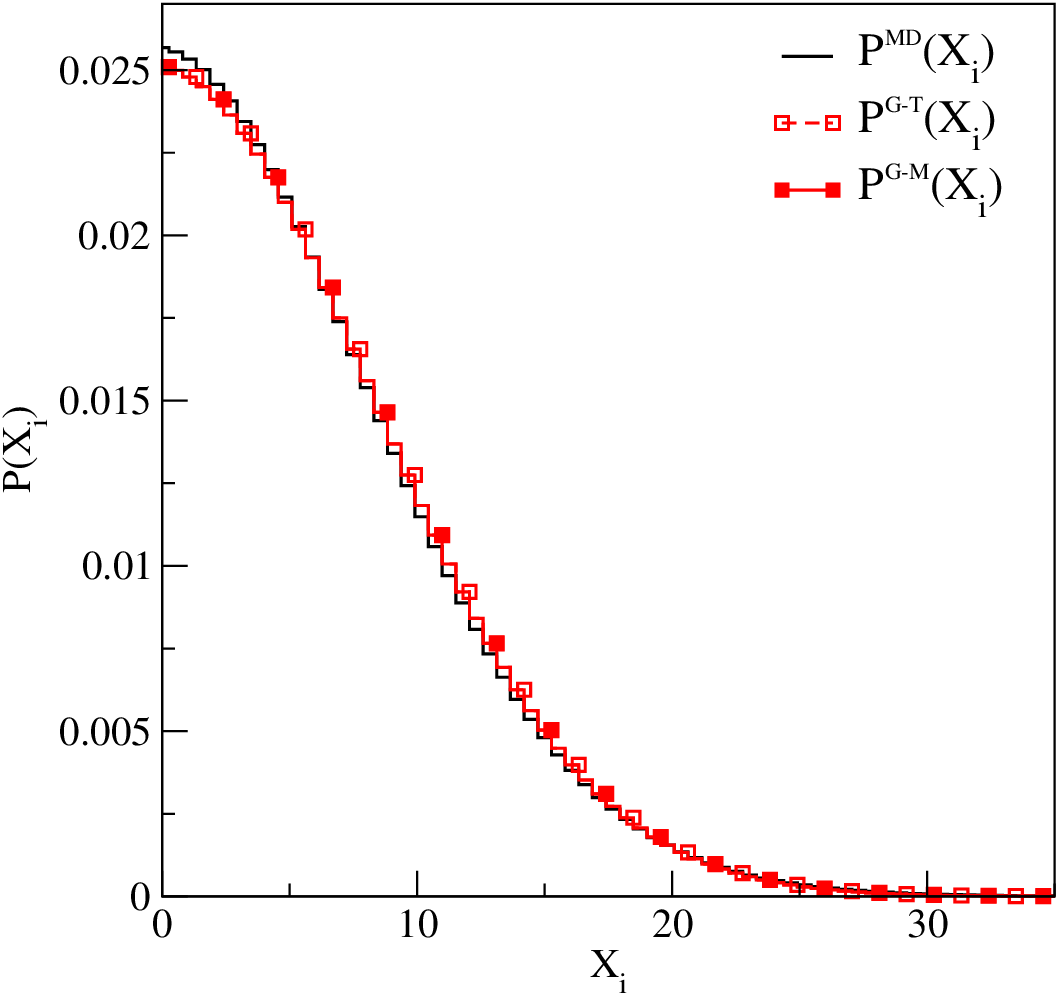}}}\\\hspace{-6mm}
     \subfloat[]{\label{subfig-2:kld_g}{\includegraphics[width=\columnwidth]{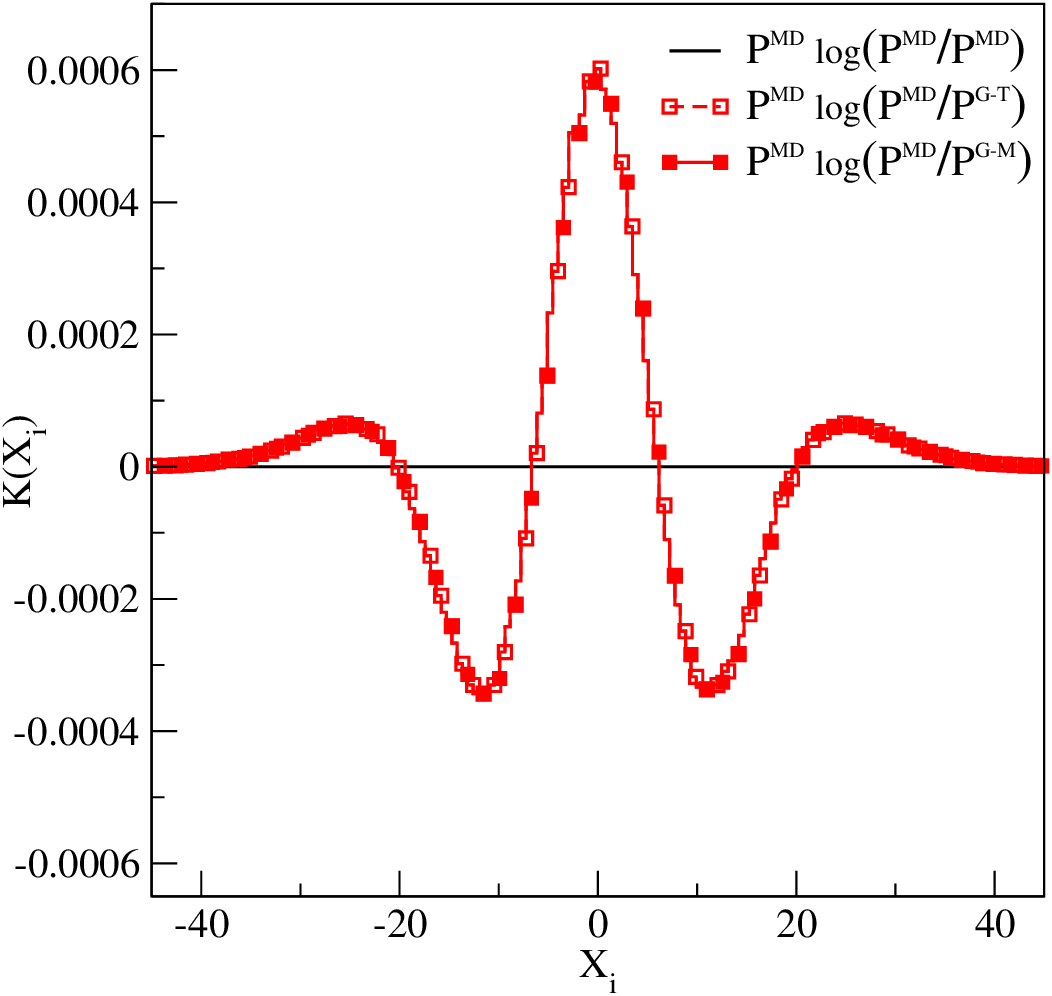}}}
     \caption{(Color online) (a) Displacements probability distribution functions. The solid line (black) depicts a PDF obtained from an MD simulation of LJ particles in equilibrium. The lines with empty or full squares (red) illustrate a Gaussian probability distribution function defined in Eq.\,(\ref{eq:p_msd}) with mean squared displacement obtained from the velocity auto-correlation function as given in Eq.\,(\ref{eq:msd}) and with mean squared displacement fitted directly to the MD data, respectively. Only the data for positive velocities has been depicted due to symmetry. (b) shows the difference between the distributions per interval $X_i$ as defined in Eq.\,(\ref{eq:k_Xi}). The presented data is for the standard parameters used in the paper and a coarse-grained time step $\Delta t = 3.2$.}
     \label{fig:kld_g}
\end{figure}

This lead us to investigate the distribution of displacements for different finite time steps. For very short time steps $\Delta t$, the effect of particle interactions can be neglected and particles simply displace according to their current velocity. Therefore, the particle displacement can be expressed as a function of the velocity and given by $\delta x_j = v_j\Delta t$ for particle $j$. The Maxwell-Boltzmann distribution function $P_v(v_j)$ as given in Eq.\,(\ref{eq:maxwell_distr}) can be expressed in terms of the particle displacements for the limiting case of $\Delta t\rightarrow 0$ as
\begin{equation}
    P(\delta x_j)=P_v\left(\frac{\delta x_j}{\Delta t}\right),
\end{equation}
and it is given by a Gaussian distribution which is fully defined by its mean value and standard deviation. Without loss of generality, we set the net momentum of our simulations to zero which corresponds to zero mean value of the distribution function. The standard deviation can be obtained in two ways: measured directly from the MD simulation; or approximated from the velocity auto-correlation function. By calculating the mean squared displacement from an analytical approximation of the velocity auto-correlation function, we obtain a simple dependence including only one parameter. Details about the performed MD simulations, the derivation of the Gaussian distribution function and discussion of the results can be found in Sections \ref{sec:setup} and \ref{sec:gaussian}. 

Regardless of the used method to obtain the mean squared displacement, Fig.\,\ref{fig:kld_g} shows that the resulting Gaussian functions -- $P^\mathrm{G{\text -}T}(X_i)$ and $P^\mathrm{G{\text -}M}(X_i)$, do not agree with the measured MD probability distribution function $P^\mathrm{MD}(X_i)$. As suspected from our studies of the deviation of non-equilibrium systems from equilibrium \cite{pachalieva_unpublished}, the equilibrium distribution functions are close to a Gaussian distribution but they show noticeable deviations from the MD data. We would like to emphasize that even though the disagreement between the two displacement functions is indeed small, it is of the same order of magnitude or larger than the deviation of a non-equilibrium distribution function.  

In this paper, we investigate for which time steps the displacement distribution is given by a Maxwell-Boltzmann distribution and when a better description is needed. We found that the Maxwell-Boltzmann function is only valid in the extreme ballistic regime for very short $\Delta t$, and in the extreme diffusive regime for very large $\Delta t$. In an intermediate regime, the Maxwellian does not capture the distribution of the displacements and introduces an error to the collision operator. This is a practical issue that matters in many meso-scale methods such as Brownian Dynamics \cite{ermak1978brownian}, Dissipative Particle Dynamics \cite{hoogerbrugge1992simulating}, Stochastic Rotation Dynamics \cite{ihle2001stochastic}, and the lattice Boltzmann method \cite{qian_lattice_1992,he1997theory}.

\section{Simulation Setup}
\label{sec:setup}
We are investigating a system of particles interacting with the standard 6-12 Lennard-Jones (LJ) intermolecular potential defined as 
\begin{equation}
    V_{LJ}(r) = 4\varepsilon\left[\left(\frac{\sigma}{r}\right)^{12}-\left(\frac{\sigma}{r}\right)^{6}\right]
    \label{eq:lj_potential}
\end{equation}
with $\varepsilon$ being the potential well depth, $\sigma$ is the distance at which the inter-particle potential goes to zero, and $r$ is the distance between two particles. We set the particle mass to $m=1$ and the LJ particle diameter to $\sigma=1$. All the MD simulations were executed using the open-source molecular dynamics software LAMMPS \cite{plimpton_fast_1995, noauthor_lammps_nodate} that is developed by Sandia National Laboratories. We performed multiple MD simulations with $N=99856$ particles in an 2D square with length $L=1000$ LJ units which corresponds to an area fraction of $\phi = 0.078387$. The area fraction $\phi$ for circular LJ particles with radius $a=\sigma/2$ is defined as the product of the particle surface area and the number of particles, divided by the square length $L$ of the simulation box. The simulations were initialised with homogeneously distributed particles having kinetic energy that corresponds to a temperature of 20 in LJ units. 

We have executed simulations of two-dimensional systems instead of three-dimensional ones to minimize computational cost. For a three-dimensional MD simulation to be computationally feasible, we need to reduce the domain size and adjust the number of particles to recover the same volume fraction as the 2D area fraction mentioned earlier. By reducing the domain size, we put a constraint on the coarse-grained time step $\Delta t$ and therefore, on the maximum average particle displacement. Thus, it will not be possible to simulate extremely large time steps due to periodic image problems occurring when the particle displacements are larger than half of the simulation length $L$.

According to the definition of the LJ interaction potential in Eq.\,(\ref{eq:lj_potential}), we write the time scale as
\begin{equation}
    \tau_\mathrm{LJ}=\sqrt{\frac{m\sigma^2}{\varepsilon}},
\end{equation}
which corresponds to the time in which a particle with kinetic energy of half the potential energy well $\varepsilon$ traverses one diameter $\sigma$ of a LJ particle. It is worth noting that there is a second time scale, i.e. the time it takes a particle with the kinetic energy of $1/2\;k_BT$ to transverse the diameter $\sigma$ of a LJ particle, which is given by 
 \begin{equation}
     \tau_\mathrm{th}=\sqrt{\frac{m\sigma^2}{k_B T}}.
 \end{equation}
and we call this scale a thermal time scale. Note that for the temperature of $20$ in LJ units, the thermal time scale is smaller than the LJ time scale $\tau_\mathrm{LJ}$ by factor of $1/\sqrt{20}\approx 0.22$.

The simulation setup characterizes a standard semi-dilute gas in equilibrium with average velocity fixed to zero
\begin{equation}
    N u_\alpha = \sum_{j=1}^N v_{j,\alpha}= 0,
    \label{eq:avg_vel}
\end{equation}
with N being the total number of MD particles.

The MD step size is set to $0.0001\,\tau_\mathrm{LJ}$ with total MD simulation time varying from $50\,\tau_\mathrm{LJ}$ to $51\,200\,\tau_\mathrm{LJ}$ as shown in Table\,\ref{tab:md_setup}. We chose a very small MD step size to ensure high accuracy of the MD simulation data. Our goal is to obtain results for MD simulations with wide regime range -- from simulations with mean free time smaller than the time between collisions (ballistic regime) to simulations with much larger mean free path than the time step (diffusive regime). We define the dimensionless coarse-grained time step $\Delta t$ as a product of the MD step size and the MD output frequency. The coarse-grained time step $\Delta t$ varies from $0.01\,\tau_\mathrm{LJ}$ to $25.6\,\tau_\mathrm{LJ}$. To ensure the MD simulations have reached equilibrium state before we start collecting data, the initial $1\,200\,000$ MD iterations ($120\,\tau_\mathrm{LJ}$) were discarded. The values depicted in Table\,\ref{tab:md_setup} do not include the discarded iterations for clarity. The total number of saved MD iterations of per particle data differs depending on the coarse-grained time step $\Delta t$. For simulations with smaller time step $\Delta t\in\interval{0.01}{0.4}$, we saved $5\,000$ coarse-grained iterations, while for simulations with $\Delta t\in\interval{0.8}{25.6}$, the output number was reduced to $2\,000$ coarse-grained iterations due to their high computational cost. This corresponds to $500\,000$ MD time steps for the MD simulation with the smallest executed coarse-grained time step $\Delta t = 0.01$, and $512\,000\,000$ MD time steps for the simulation with the largest time step $\Delta t = 25.6$. Since we are simulating a semi-dilute gas in equilibrium, the total simulation number is irrelevant for the physical properties of the system, because they do not change once the gas has reached equilibrium state. However, we run the simulations for large number of iterations in order to produce large amounts of data which ensures sufficient averaging. An overview of the simulation parameters is given in Table\,\ref{tab:md_setup}.
\begin{table} 
\caption{LAMMPS Simulation Details.
\label{tab:md_setup}}
\begin{ruledtabular}
\begin{tabular}{  c | c | c | c | c | c }
            & MD  step           & MD output  & Output      & Total MD     & Total MD  \\
$\Delta t $ & size ($\tau_\mathrm{LJ}$) & frequency  &  number     &  time steps  & time ($\tau_\mathrm{LJ}$)   \\ 
\hline 
 0.01 & 0.0001  &      100   & 5000     &      500\,000   &       50    \\
 0.1  & 0.0001  &   1\,000   & 5000     &   5\,000\,000   &      500    \\
 0.2  & 0.0001  &   2\,000   & 5000     &  10\,000\,000   &   1\,000    \\
 0.4  & 0.0001  &   4\,000   & 5000     &  20\,000\,000   &   2\,000    \\
 0.8  & 0.0001  &   8\,000   & 2000     &  16\,000\,000   &   1\,600    \\
 1.6  & 0.0001  &  16\,000   & 2000     &  32\,000\,000   &   3\,200    \\
 3.2  & 0.0001  &  32\,000   & 2000     &  64\,000\,000   &   6\,400    \\
 6.4  & 0.0001  &  64\,000   & 2000     & 128\,000\,000   &  12\,800    \\
12.8  & 0.0001  & 128\,000   & 2000     & 256\,000\,000   &  25\,600    \\
25.6  & 0.0001  & 256\,000   & 2000     & 512\,000\,000   &  51\,200    \\
\end{tabular}
\end{ruledtabular}
\end{table}

All the simulations were executed in parallel using 32 processors on the Darwin cluster at Los Alamos National Laboratory (LANL). The longest executed test case with $\Delta t = 25.6$ took about 120 hours wall-clock-time. Depending on the number of coarse-grained iterations ($2\,000$ or $5\,000$) the output data files took 20\,GB or 50\,GB memory space, respectively. The total memory space used for all LAMMPS simulations exceeded 350\,GB.

We have performed constant NVE integration to update atoms' position and velocity each time step. By using the NVE thermostat, we sample from the microcanonical ensamble, thus we avoid any possible complications coming from the altered equations of motion a thermostat could introduce. However, to ensure the validity of the MD simulations, we have tested the canonical NVT thermostat which was used in earlier papers \cite{parsa_lattice_2017,parsa_validity_2019} and we have obtained equivalent results. Nevertheless, the results presented in the current publication are obtained using the NVE microcanonical ensamble.

We analyse the collected MD data to recover the Probability Distribution Function (PDF) $P(\delta x)$ of the displacements $\delta x$. To obtain an estimate for $P(\delta x)$, we define the particle displacement $\delta x_j(t)$ as
\begin{equation}
    \delta x_{j,\alpha}(t+\Delta t) = x_{j,\alpha}(t) - x_{j,\alpha}(t+\Delta t),
\end{equation}
where $x_{j,\alpha}(t)$ is the position of particle $j$ at time $t$, and $\alpha$ refers to the spatial coordinates $\alpha \in \{X, Y\}$ in 2D. 

Two probability distribution functions can be compared in different ways: in principle the PDF is defined as a function or it can be defined through an infinite set of moments. 
Given the experimental data set, we are of course limited in how well we can estimate the PDF. Therefore, here we use a combination of both approaches. 

To obtain the full PDF description, we define a histogram $H(X_i)$ for the discrete displacement intervals $X_i$ as follows
\begin{equation}
    H(X_i) = \frac{\sum_{t=0}^T \sum_{j=1}^N \Delta_{X_i}(\delta x_j(t))}{TN},
    \label{eq:p_xi}
\end{equation}
with number of MD particles $N$, number of the coarse-grained time steps $T$ and with $\Delta_{X_i}(\delta x_j(t))$ being defined as
\begin{equation}
    \Delta_{X_i}(\delta x_j(t)) = \left\{\begin{array}{ll}
    1, & \mbox{if } \delta x_j(t)\in X_i\\
    0, & \mbox{otherwise.}
    \end{array}\right.
\end{equation}
$X_i$ is a histogram bin and corresponds to a range of $r_i \leq \delta x < r_{i+1}$ with $i$ number of bins. In the current publication, we use $i=200$ number of bins with equal bin width for a certain coarse-grained time step. The bin width depends on the particle displacements and varies for different time step $\Delta t$. The first and the last intervals are open at the edges to ensure that there are no empty bins in the histogram and that all possible displacements have been accounted for. This histogram has the following property
\begin{equation}
    \sum_i H(X_i)=1.
\end{equation}
We can then estimate the probability
\begin{equation}
    P(\delta x\in X_i)=\int_{\delta x\in X_i} P(\delta x)\,d\delta x \approx H(X_i).
    \label{eq:p_Hi}
\end{equation}
Even though the MD data is in discrete space and by using the collected MD displacements we are able to construct only a histogram as given in Eq.\,(\ref{eq:p_xi}), we will further recall it as a probability distribution function. By collecting very large data sets for each coarse-grained time step $\Delta t$, we ensure that all histograms are very fine grained and thus agree very well with the underlying PDF as expressed in Eq.\,(\ref{eq:p_Hi}).

In our MD simulation setup, momentum is conserved. This means that we can also define the momentum through the displacements in addition to Eq.\,(\ref{eq:avg_vel}). We have
\begin{equation}
    u_\alpha = \frac{\langle \delta x_{j,\alpha} \rangle}{\Delta t}
    =\frac{\sum_{j=1}^N \delta x_{j,\alpha}}{N \Delta t}
    =\frac{\sum_{j=1}^N v_{j,\alpha}}{N},
    \label{eq:mean_vel}
\end{equation}
which are all equivalent. Even though, we have performed simulations with zero initial velocity we could obtain results for different mean velocities $u_{\alpha}$ by applying a Galilean transformation.

\section{Gaussian Distribution Function}
\label{sec:gaussian}
\begin{figure}[!htbp]
     \centering
     \subfloat[]{\label{subfig-1:vel_corr}{\includegraphics[width=\columnwidth]{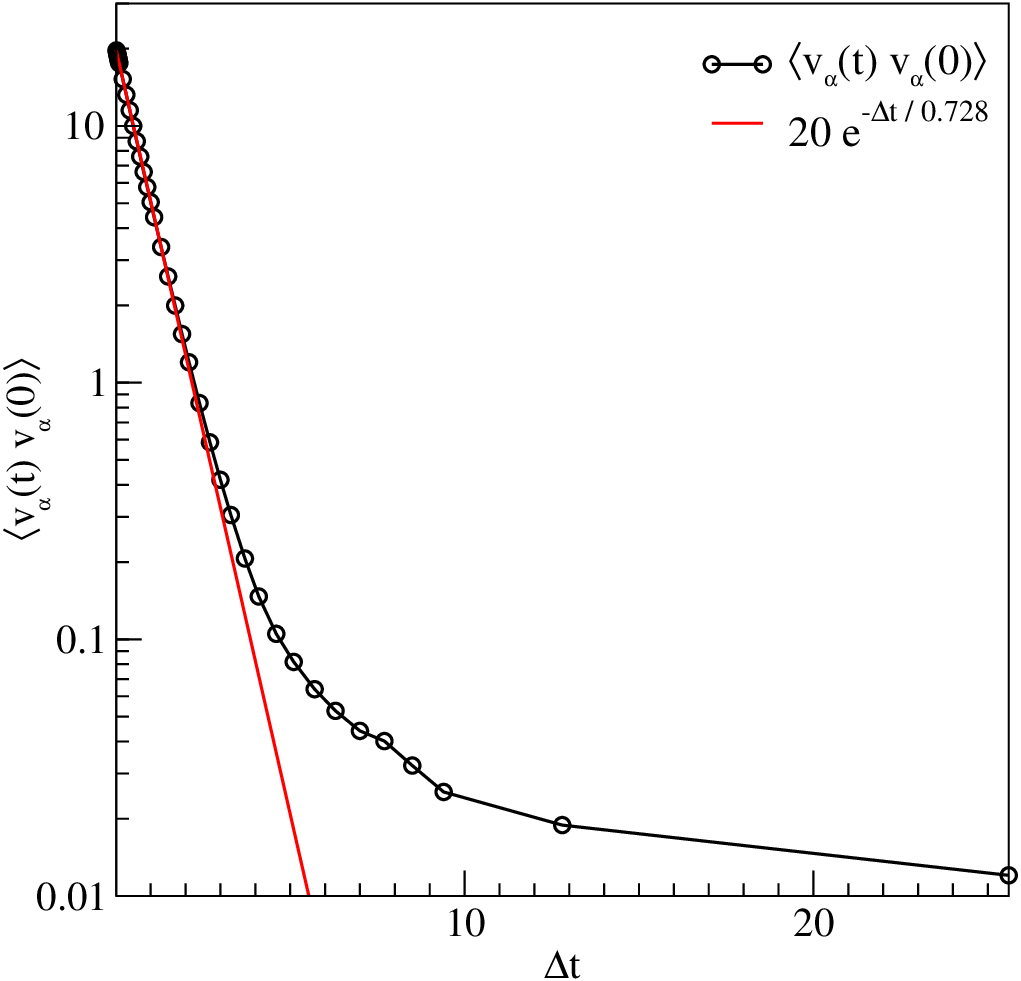}}}\\\hspace{-6mm}
     \subfloat[]{\label{subfig-2:msd}{\includegraphics[width=\columnwidth]{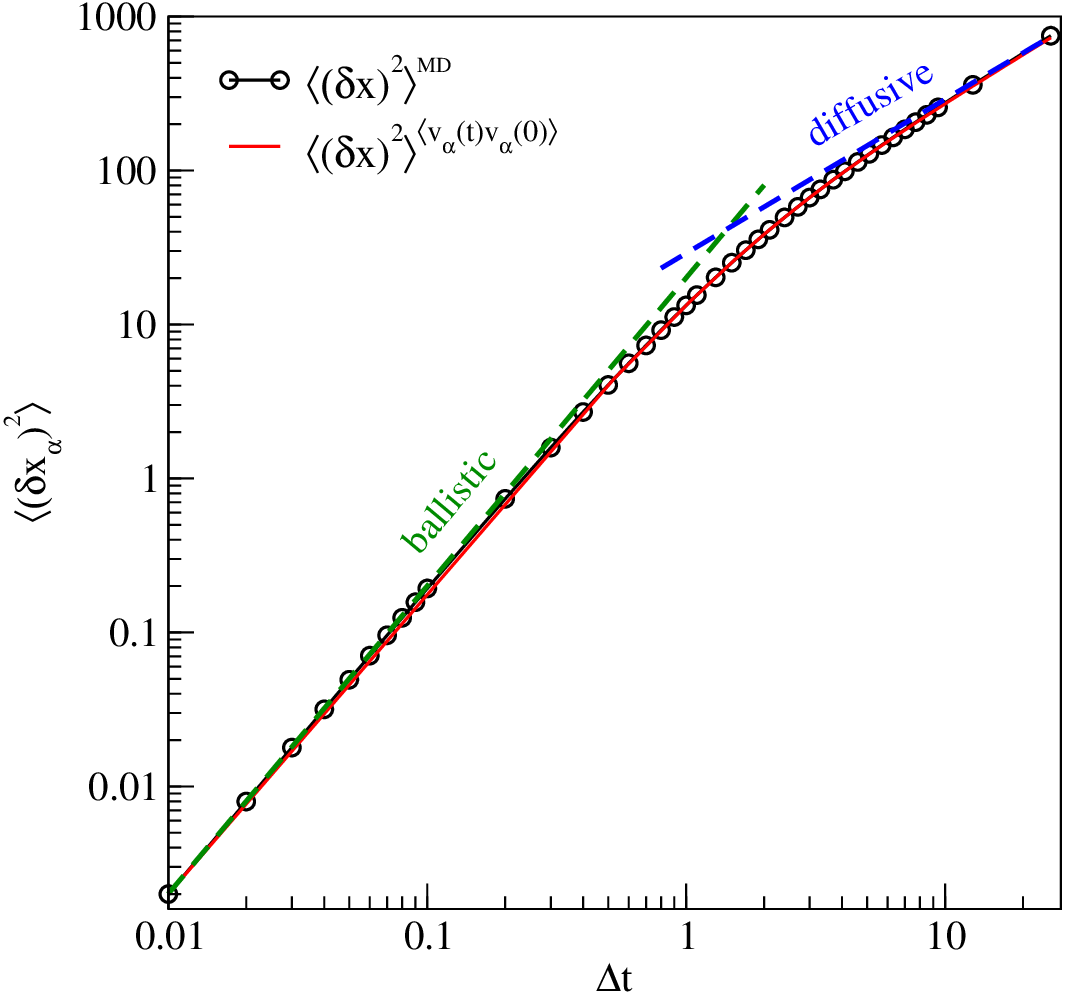}}}
     \caption{(Color online) (a) Velocity auto-correlation function measured from an MD simulation compared to an exponential decay with $\tau\approx 0.728$ as given in Eq.\,(\ref{eq:vel_corr_exp_decay}). The long time tails are typical for two-dimensional systems \cite{uhlenbeck_theory_1930, green_markoff_1952, green_markoff_1954,kubo_ryogo_statistical-mechanical_1957,weitz_nondiffusive_1989} (b) The mean squared displacement directly measured from an MD simulation is compared to the theoretical value given in Eq.\,(\ref{eq:msd}). Notice the two scaling regimes: $\langle(\delta x_\alpha)^2\rangle\propto\Delta t$ for a ballistic regime with small times; and $\langle(\delta x_\alpha)^2\rangle\propto\Delta t^2$ for a diffusive regime with large times.}
     \label{fig:vel_corr_msd}
\end{figure}

The first theory for the probability distribution function of the displacements that we consider follows the assumption made by Parsa \textit{et al.} \cite{parsa_lattice_2017}. For very short times the particle displacement is given by the velocity $v_j$ of the particle $j$ as $\delta x_j = v_j \Delta t$. Thus, we can write $\lim_{\Delta t\rightarrow 0} P(\delta x_j) = P_v(\delta x_j/\Delta t)$ using the probability distribution of the velocity given by
\begin{equation}
    P_v(v_j)=\frac{1}{[2\pi k_BT]^{d/2}}\exp\left(\frac{(v_j-u_j)^2}{2k_BT}\right),
    \label{eq:maxwell_distr}
\end{equation}
where $d$ is the number of dimensions, $k_BT$ is temperature of the system with $k_B$ being the Boltzmann constant. Eq.\,(\ref{eq:maxwell_distr}) is also known as the Maxwell-Boltzmann distribution which approximates the probability of particle moving in a certain direction. It holds for very short times $\Delta t$ where the mean free time between two collisions is much shorter than the time step $\Delta t$. In this regime, particles undergo simple ballistic motion and the mean squared displacement in one dimension is
\begin{equation}
    \langle(\delta x_\alpha)^2\rangle^\mathrm{ball} = 2k_BT(\Delta t)^2.
    \label{eq:ball_sd}
\end{equation}
Then the probability for collisionless displacements is
\begin{equation}
   P^\mathrm{ball}(\delta x)=\frac{1}{[2\pi k_BT(\Delta t)^2]^{d/2}}\exp\left(-\frac{(\delta x-u\Delta t)^2}{2 k_BT(\Delta t)^2}\right).
    \label{eq:p_ball}
\end{equation}
In a diffusive regime, the times are much longer than the mean free time and the particles undergo multiple collisions between time steps. Using the self-diffusion constant $D$, we write the mean squared displacement in one dimension as
\begin{equation}
    \langle(\delta x_\alpha)^2\rangle^\mathrm{diff}= 2d D(\Delta t).
    \label{eq:diff_sd}
\end{equation}
The probability distribution function of the displacements is given by 
\begin{equation}
    P^\mathrm{diff}(\delta x)=\frac{1}{[4\pi d D(\Delta t)]^{d/2}}\exp\left(-\frac{(\delta x-u\Delta t)^2}{4 d D(\Delta t)}\right).
    \label{eq:p_diff}
\end{equation}
Since both limiting cases are given by a Gaussian distribution function as shown in Eqs.\,(\ref{eq:p_ball}) and (\ref{eq:p_diff}), Parsa \textit{et al.} \cite{parsa_lattice_2017} suggested that the intermediate probabilities can be well approximated by a single Gaussian distribution defined as 
\begin{equation}
    P^\mathrm{G}(\delta x)=\frac{1}{[2\pi \langle(\delta x_\alpha)^2\rangle]^{d/2}}\exp\left(-\frac{(\delta x-u\Delta t)^2}{2\langle(\delta x_\alpha)^2\rangle}\right),
    \label{eq:p_msd}
\end{equation}
with a mean squared displacement $\langle(\delta x_\alpha)^2\rangle$ which can be obtained theoretically or can be measured directly from an MD simulation. The displacement of a particle is given by
\begin{equation}
    \delta x = \int_0^{\Delta t} v(t)\,dt.
    \label{eq:def_displ}
\end{equation}
Now, for a simple semi-dilute gas system, we express the mean squared displacement as a function of the velocity auto-correlation function
\begin{equation}
\begin{split}
    \langle(\delta x_\alpha)^2 \rangle &=\left\langle \int dt \int dt' v(t) v(t')\right\rangle \\
    &=\int dt \int dt' \left\langle v(t-t') v(0)\right\rangle\\
    &= \int_{-\Delta t}^{\Delta t} (\Delta t-\delta t)\langle v(\delta t) v(0)\rangle\,d\delta t\\
    &= 2 \int_{0}^{\Delta t} (\Delta t-\delta t)\langle v(\delta t) v(0)\rangle\,d\delta t
    \label{eq:def_vcf}
\end{split}
\end{equation}
For gases the velocity auto-correlation function is usually estimated by an exponential decay
\begin{equation}
    \langle v_\alpha(\delta t)v_\alpha(0)\rangle = k_BT\exp\left(-\frac{\Delta t}{\tau}\right),
    \label{eq:vel_corr_exp_decay}
\end{equation}
where $k_BT$ is the temperature of the semi-dilute gas in LJ units, and $\tau$ is an exponential decay constant which approximates the mean free time \cite{uhlenbeck_theory_1930, green_markoff_1952, green_markoff_1954,kubo_ryogo_statistical-mechanical_1957,weitz_nondiffusive_1989}. The velocity auto-correlation function for the simulated gas system is depicted in Fig.\,\ref{subfig-1:vel_corr}. We have approximated the mean free time to $\tau\approx0.728$, which gives a good prediction of the velocity auto-correlation function for early times. As shown in Fig.\,\ref{subfig-1:vel_corr}, the velocity auto-correlation function has long range contributions for later times ($\Delta t>4.0$) that is typical for two-dimensional systems \cite{uhlenbeck_theory_1930, green_markoff_1952, green_markoff_1954,kubo_ryogo_statistical-mechanical_1957,weitz_nondiffusive_1989}. The deviations resulting from the long time tails are noticeable only for later times and larger displacements. In this work, we focus on results for $\Delta t=3.2$, where the velocity auto-correlation function is well approximated by an exponential decay as defined in Eq.\,(\ref{eq:vel_corr_exp_decay}). For simplicity, we will therefore neglect the long time tails shown in Fig.\,\ref{subfig-1:vel_corr}.

Now, the theoretical mean squared displacement can be calculated according to Eq.\,(\ref{eq:def_vcf}) as 
\begin{equation}
    \langle(\delta x_\alpha)^2\rangle = 2k_BT\tau^2\left(\exp\left(-\frac{\Delta t}{\tau}\right)+\frac{\Delta t}{\tau}-1\right).
    \label{eq:msd}
\end{equation}
As shown in Fig.\,\ref{subfig-2:msd}, this prediction recovers the mean squared displacement very well. There are small deviations for later times which are not visible in log-log scale. These deviations are result of the long time tails of the velocity auto-correlation function mentioned previously. This completes the definition of the Gaussian distribution function model using a mean squared displacement obtained from Eq.\,(\ref{eq:msd}). In general, $\langle(\delta x_\alpha)^2\rangle$ can be also measured from the MD simulations. Later, we compare the Gaussian distribution functions obtained using these two approaches.

To estimate how good this PDF matches the MD data, we transform the formulation of $P(\delta x)$ from continuous to discrete using a histogram as defined in Eq.\,(\ref{eq:p_Hi}). This is realized  by integrating the probability distribution function over predefined intervals $X_i$ as
\begin{equation}
    \begin{split}
        H(X_i)&=\int_{r_{i}}^{{r_{i+1}}}P(\delta x)\,d\delta x\\ 
        &= \frac{1}{2}\left[\mathrm{erf}\left(\frac{r_{i}}{\sigma\sqrt{2}}\right)-\mathrm{erf}\left(\frac{r_{i+1}}{\sigma\sqrt{2}}\right)\right]
    \end{split}
\end{equation}
where $X_i$ corresponds to a rage of $r_i \leq \delta x < r_{i+1}$ with number of bins $i=200$. $\mathrm{erf}(r_i)$ is an error function encountered in integrating the normal distribution function with standard deviation $\sigma$ and mean equal to zero. Using a histogram to compare two PDFs is a convenient method to analyze precisely where two or more distributing functions diverge.


To analyse how well the Gaussian distribution function fits the MD displacements in the transition regime, we consider a time step of $\Delta t=3.2$. In Fig.\,\ref{subfig-1:df_g}, the MD displacements (black line) are plotted alongside a Gaussian distribution function $P^\mathrm{G{\text -}T}(X_i)$ with theoretical mean squared displacement (red dashed line with empty squares) and a Gaussian distribution function $P^\mathrm{G{\text -}M}(X_i)$ with measured mean squared displacement (red line with full squares). Both Gaussian distribution functions give an adequate prediction of the MD displacements distribution function, however, there are visible discrepancies at about 5\%. Even though the deviations between the MD data and the proposed Gaussian distribution functions are small, they are of significant importance when examining non-equilibrium behavior and looking at small deviations from equilibrium. 

Since the deviations between the Gaussian PDFs and the MD simulation data are relatively small, the following function is used to quantify more precisely the discrepancies
\begin{equation}
     K(X_i) = K(R\parallel Q) = R(X_i)\log \left({\frac {R(X_i)}{Q(X_i)}}\right)
     \label{eq:k_Xi}
\end{equation}
where $R(X_i)$ and $Q(X_i)$ are probability distributions over an interval $X_i$. By performing a sum over all the bins $X_i$, we obtain the well known Kullback-Leibler (KL) divergence \cite{kullback1951information} defined as 
\begin{equation}
     D_{\text{KL}}(R\parallel Q)=\sum_{i}R(X_i)\log \left({\frac {R(X_i)}{Q(X_i)}}\right).
     \label{eq:kullback_leibler}
\end{equation}
The KL divergence measures the discrepancies of one probability distribution function to another. It is always non-negative $D_{\text{KL}}(R\parallel Q)\geq0$ or equal to zero if and only if the probability distribution functions are identical $R(X_i) = Q(X_i)$ \cite{kullback1951information}. 

In Fig.\,\ref{subfig-2:kld_g}, we show the discrepancies between the Gaussian probability distribution functions and the MD data per bin element $X_i$ measured using Eq.\,(\ref{eq:k_Xi}). The solid line (black) depicts $K(P^\mathrm{MD}\parallel P^\mathrm{MD})$ which is zero by construction. The lines with full or empty symbols (red) display the divergence between the MD data and the Gaussian distribution functions with theoretical or measured mean squared displacement, respectively. Note here that the $K(X_i)$ measure identifies both positive and negative deviations (which is necessary, since the integral of both probability distribution functions is 1) but as long as there is any deviation, the integral (or sum) in Eq.\,(\ref{eq:kullback_leibler}) always leads to a positive value.  We can see a clear structure in the error of the MD data and the two Gaussian probability distribution functions. Thus, we conclude that a single Gaussian distribution function with the same standard deviation, being measured or theoretically obtained from the velocity auto-correlation function, differs significantly from the MD data in the intermediate regime.

\begin{figure}[!tbp]
     \centering
     \includegraphics[width=\columnwidth]{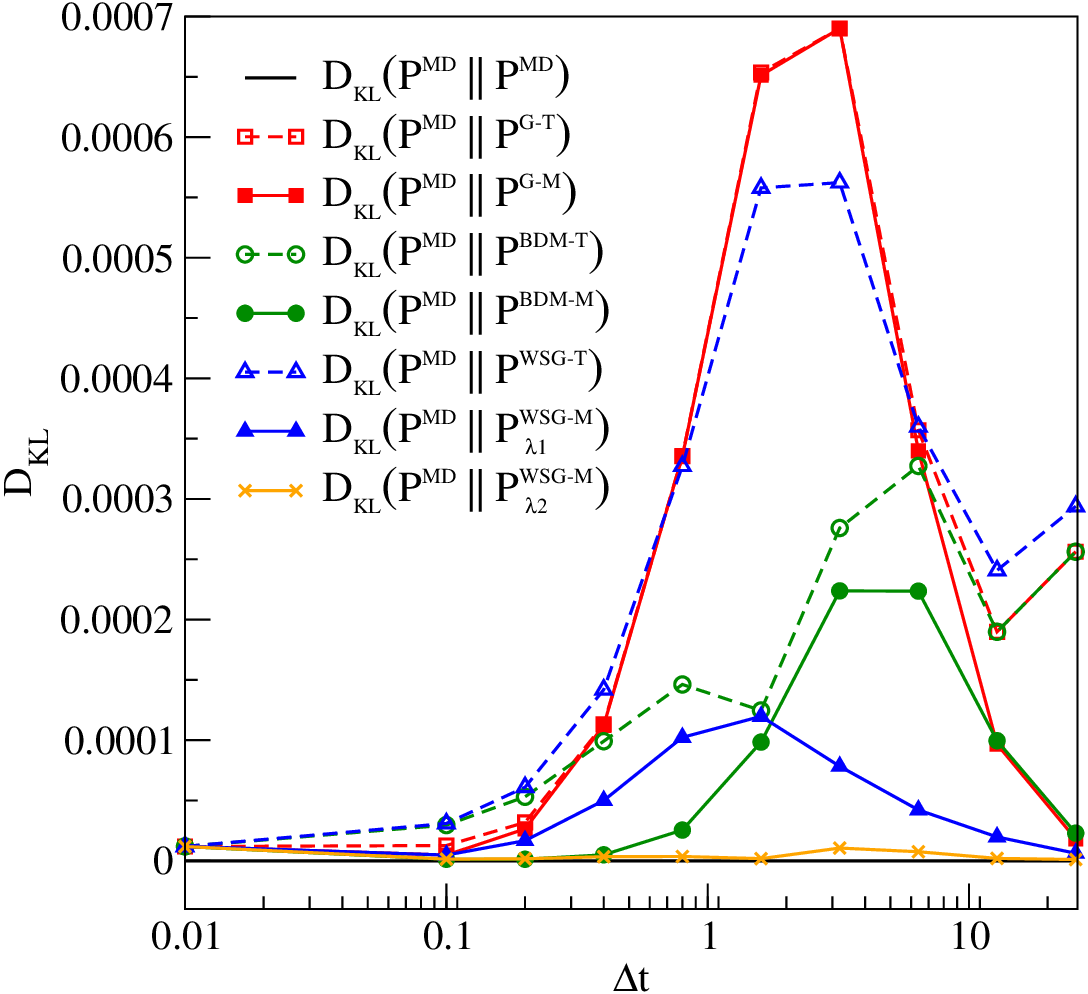}
     \caption{(Color online) Kullback-Leibler divergence results: empty/full squares (red) for $D_\mathrm{KL}(P^\mathrm{MD}\parallel P^\mathrm{G{\text -}T})$ and $D_\mathrm{KL}(P^\mathrm{MD}\parallel P^\mathrm{G{\text -}M})$ discussed in Section\,\ref{sec:gaussian}; empty/full circles (green) for $D_\mathrm{KL}(P^\mathrm{MD}\parallel P^\mathrm{BDM{\text -}T})$ and $D_\mathrm{KL}(P^\mathrm{MD}\parallel P^\mathrm{BDM{\text -}M})$ discussed in Section\,\ref{sec:ball_diff}; empty/full triangles (blue) for $D_\mathrm{KL}(P^\mathrm{MD}\parallel P^\mathrm{WSG{\text -}T})$ and $D_\mathrm{KL}(P^\mathrm{MD}\parallel P^\mathrm{WSG{\text -}\lambda_1})$, and x-symbols (yellow) for $D_\mathrm{KL}(P^\mathrm{MD}\parallel P^\mathrm{WSG{\text -}\lambda_2})$ discussed in Section\,\ref{sec:poisson_gaussians}. The $D_\mathrm{KL}(P^\mathrm{MD}\parallel P^\mathrm{MD})$ divergence (black line) is zero by definition and it is shown just as a comparison. All displacements PDFs show small error for very small $\Delta t$ (ballistic regime) and for large $\Delta t$ (diffusive regime). However, in the transition regime only the $P^\mathrm{WSG{\text -}\lambda_2}(X_i)$ distribution function with average number of collisions $\lambda_2$ gives a satisfactory description of the measured MD distribution function. The KL divergence is calculated for all time steps $\Delta t\in\interval{0.01}{25.6}$ considered in this publication.}
     \label{fig:kld_error}
\end{figure}

The Kullback-Leibler divergence of the PDF models and the MD data is illustrated in Fig.\,\ref{fig:kld_error}. The divergence is calculated for a variety of time steps $\Delta t \in [0.01,25.6]$. In the current section, we focus on the two KL divergence measures $D_\mathrm{KL}(P^\mathrm{MD}\parallel P^\mathrm{G{\text -}T})$ and $D_\mathrm{KL}(P^\mathrm{MD}\parallel P^\mathrm{G{\text -}M})$ depicted by lines with full or empty squares (red), respectively. As expected, for purely ballistic test cases the constructed Gaussian distribution functions match very well the PDF obtained from MD data. In the transition regime, the estimated divergence increases rapidly and reaches a peak at $\Delta t=3.2$, which indicates that the MD displacement function cannot be captured using a single Gaussian distribution function. For $\Delta t=25.6$, the $K(P^\mathrm{MD}\parallel P^\mathrm{G{\text -}M})$ divergence is close to zero and we conclude that the simulation has reached diffusive regime. For some of the considered time steps, $P^\mathrm{G{\text-}M}(X_i)$ delivers slightly better results in comparison to $P^\mathrm{G{\text-}T}(X_i)$ but the improvement is not significant. For the particular case of $\Delta t=3.2$, there is no visible difference between the two Gaussian distribution functions, which explains the complete overlap of the $K(P^\mathrm{MD}\parallel P^\mathrm{G{\text -}T})$ and $K(P^\mathrm{MD}\parallel P^\mathrm{G{\text -}M})$ results shown in Fig.\,\ref{subfig-2:kld_g}.

To obtain a better theoretical formulation for the distribution of the equilibrium LJ displacements, we need to analyze rigorously the displacements' distribution function obtained from the MD data. One way to distinguish between two distribution functions is by looking at their moments. By estimating the PDF using the moments of the MD displacements, we eliminate the small error introduced by the histogram in Eq.\,(\ref{eq:p_Hi}). From the MD simulation data, we calculate the $k^\mathrm{th}$ moment as
\begin{equation}
    \mu_k = \langle (\delta x)^k \rangle.
\end{equation}
Since we are looking at an ensemble average of particle displacements, the moments $\mu_k$ can be averaged in space and in time, leading to the following approximation 
\begin{equation}
 \mu_k = \frac{\sum_{t=1}^T \sum_{j=1}^N (x_j(t+\Delta t)-x_j(t))^k}{TN}
\end{equation}
with $N$ being the number of MD particles and $T$ being the number of the coarse-grained time steps. The zeroth moment is given simply by the normalization as $\mu_0=1$. The first moment defines the average velocity $u_\alpha$, which in our simulation setup is zero and leads to zero first and third order moments $\mu_1 = \mu_3 = 0$ due to symmetry. The second moment $\mu_2$ is known in statistics as the variance or the mean squared displacement and is given by $\mu_2 = \langle(\delta x)^2\rangle$. The fourth moment $\mu_4 = \langle(\delta x)^4\rangle$ is called kurtosis and it is a measure for the "tailedness" of a probability distribution function. 

A probability distribution function is defined uniquely through an infinite set of moments. Generally, the better moments match, the better the distributions agree, and the higher order a moment is the less important it tends to be. It is therefore reasonable that we examine the agreements of the moments. The zeroth moment corresponds to normalization and always matches. The second moment should always match, but small errors can occur for theoretical distributions that use Eq.\,(\ref{eq:msd}). The fourth order moments at this point are unconstrained, and therefore the deviation of this moment from the experimental one should give a good estimate of the accuracy of the theoretical distribution. We therefore focus on the first two nontrivial moments -- $\mu_2$ and $\mu_4$. The moments $\mu_0, \mu_1$ and $\mu_3$ have been measured for completeness, but their value for LJ particles in equilibrium are expected to be $\mu_0=1$, and $\mu_1=\mu_3=0$ for symmetry reasons. 

\begin{figure}[!tbp]
     \centering
     \includegraphics[width=\columnwidth]{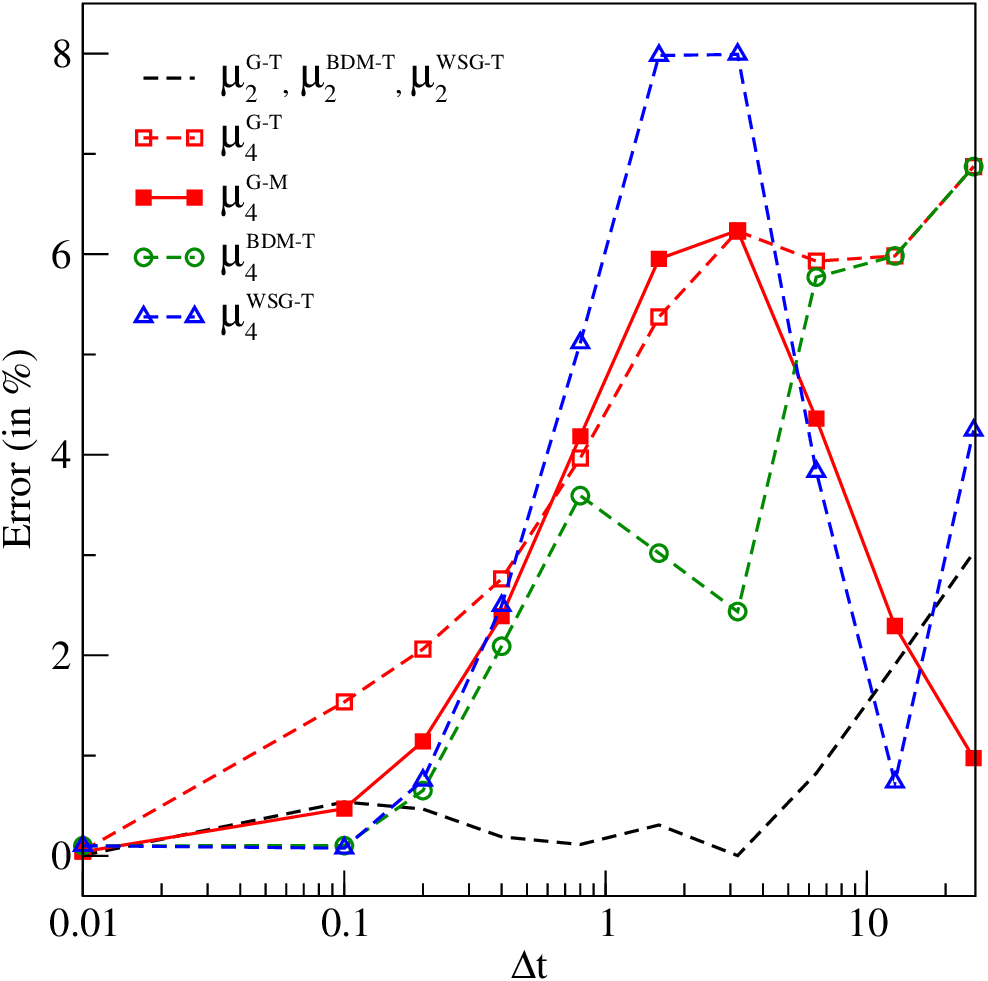}
     \caption{(Color online) Second and fourth order moment error calculated between the MD simulation data and the theoretical probability distribution functions. The second order error is equivalent for all theoretical PDF models. The fourth order error varies: the $P^\mathrm{G{\text -}T}(\delta x)$ and $P^\mathrm{G{\text -}M}(\delta x)$ errors are discussed in Section\,\ref{sec:gaussian} (red lines with empty/full squares); the $P^\mathrm{BDM{\text -}T}(\delta x)$ errors discussed in Section\,\ref{sec:ball_diff} (green lines with circles); and the $P^\mathrm{WSG{\text -}T}(\delta x)$ error is discussed in Section\,\ref{sec:poisson_gaussians} (blue lines with triangles). For some of the proposed distribution functions the second and the fourth order moments have been fitted to the measured MD moments. These PDFs have zero second and fourth order error by construction, therefore, they have not been depicted. The presented data is for the standard parameters used in the paper and a time step $\Delta t = 3.2$.}
     \label{fig:mu_error}
\end{figure}

As mentioned previously, the probability distribution function $P^\mathrm{G}(\delta x)$ in Eq.\,(\ref{eq:p_msd}) could be calculated using a theoretical or a measured $\langle(\delta x)^2\rangle$. We measured the second and fourth moments of the Gaussian distribution functions and compared their deviation from the MD moments as shown in Fig.\,\ref{fig:mu_error}. The error is calculated in percentage.

The Gaussian distribution function with theoretical mean squared displacement fails to reconstruct the second and the fourth order moments. The second order moment error, depicted with a dashed line (black), is relatively small (below 3\%). This error rapidly increases with larger time steps and reaches its highest point at $\Delta t=25.6$. The $P^\mathrm{G{\text -}T}(\delta x)$ fourth order moment error is depicted in Fig.\,\ref{fig:mu_error} as dashed line (red) with empty squares. The $\mu^\mathrm{G{\text -}T}_4$ error is much larger than the $\mu^\mathrm{G{\text -}T}_2$ error and increases very fast in the transition regime.

The second order moment of $P^\mathrm{G-M}(X_i)$ matches the MD second order moment by construction. The fourth order moment $\mu_4^\mathrm{G{\text -}M}$, however, differs from the measured fourth order moment as shown in Fig.\,\ref{fig:mu_error} (red line with full squares). For $\Delta t\in[0.8,1.6]$, $P^\mathrm{G{\text -}M}(\delta x)$ has a slightly larger fourth order moment error than $P^\mathrm{G{\text -}T}(\delta x)$. Unlike $\mu_4^\mathrm{G{\text -}T}$, which does not decrease with larger time steps, the $\mu_4^\mathrm{G{\text -}M}$ error is large in the transition regime and decreases to less than 1\% for later times. We assume that the larger $\mu_4^\mathrm{G{\text -}T}$ error is related to the larger second order moment error of $P^\mathrm{G{\text -}T}(\delta x)$. Fig.\,\ref{fig:mu_error} shows that the $\mu_4^\mathrm{G{\text -}M}$ error is very small for early and late times which indicates that the Gaussian description with measured mean squared displacement is valid for extreme ballistic and diffusive regimes.

Considering these results, we conclude that a single Gaussian distribution function cannot recover the MD displacements distribution function in a transition regime. In the following section, we construct a Gaussian mixture model which can be adjusted to capture better the MD simulation data.


\section{Ballistic-Diffusive Distribution Function}
\label{sec:ball_diff}
Going back to the assumption made by Parsa \textit{et al.} \cite{parsa_lattice_2017}, we take a slightly different approach by approximating the displacements PDF using a Gaussian mixture model with two components. The first component is a distribution function in a ballistic regime given by Eq.\,(\ref{eq:p_ball}), while the second component is a distribution function in a diffusive regime defined in Eq.\,(\ref{eq:p_diff}). We call this formulation Ballistic-Diffusive Mixture (BDM) model and define it as
\begin{equation}
\begin{split}
     \hspace{-3mm}P^{\mathrm{BDM}}(\delta x) &=\exp\left(-\frac{\Delta t}{\tau}\right)\,P^\mathrm{ball}(\delta x)
     \\&+\left[1-\exp\left(-\frac{\Delta t}{\tau}\right)\right]\,P^\mathrm{diff}(\delta x),
\end{split}
    \label{eq:p_bdm}
\end{equation}
where the ratio $\Delta t/\tau$ relates to the average number of collisions within a time interval $\Delta t$. The mean free time $\tau$ can be evaluated from the velocity auto-correlation function as given in Eq.\,(\ref{eq:vel_corr_exp_decay}). As shown in Fig.\,\ref{subfig-1:vel_corr}, the mean free time is estimated to $\tau\approx0.728$ which agrees well with the measured velocity auto-correlation function for early times.

In a transition regime, the BDM model receives contributions from the ballistic and from the diffusive Gaussian distribution functions. The mixing coefficient $\exp\left(-\Delta t/\tau\right)$ depends on the time step and controls the ratio of the two probability distribution functions. For infinite small or infinite large time steps, $P^\mathrm{BDM}(\delta x)$ is reduced to a single Gaussian distribution given by Eq.\,(\ref{eq:p_ball}) or Eq.\,(\ref{eq:p_diff}), respectively.

\begin{figure}[!t]
\includegraphics[width=\columnwidth]{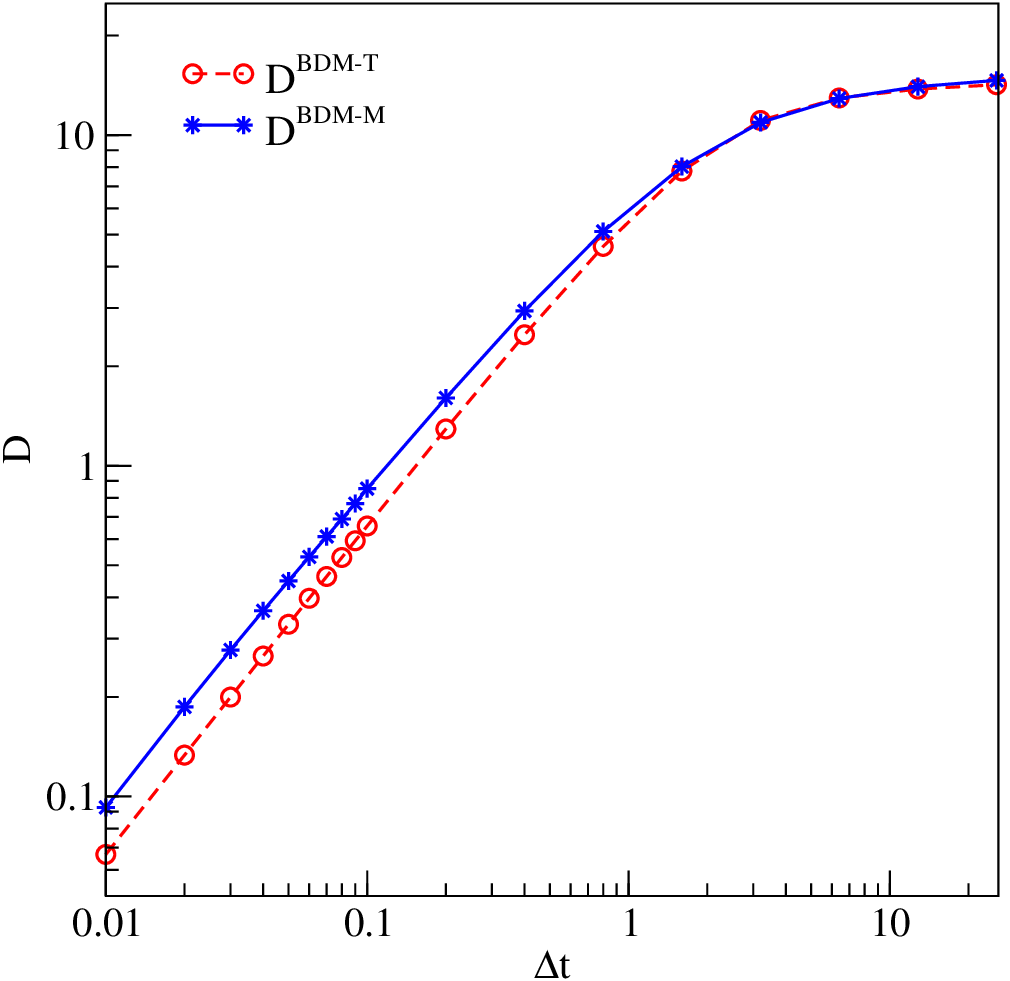}%
\caption{(Color online) Dependence of the self-diffusion constant to the time step $\Delta t$. The diffusion $D=\sigma^2_\mathrm{diff}/2\Delta t$ converges to a constant for $P^\mathrm{BDM{\text -}T}$ and $P^\mathrm{BDM{\text -}M}$, however, for early times it is not fixed. This demonstrates that the BDM models do not capture the hydrodynamics properties.}
\label{fig:diffusion}
\end{figure}

For the BDM model in Eq.\,(\ref{eq:p_bdm}), the ballistic contribution is fully defined by the simulation setup with standard deviation equal to $2k_BT(\Delta t)^2$ as given in Eq.\,(\ref{eq:ball_sd}). For the diffusive part $P^\mathrm{diff}(\delta x)$, one could attempt to simply relate it to the self-diffusion constant $D$. This does not give the correct second order moment though. Instead, we generalize the diffusive PDF from Eq.\,(\ref{eq:p_diff}) as 
\begin{equation}
    P^\mathrm{diff}(\delta x)=\frac{1}{[2\pi \sigma_\mathrm{diff}^2]^{d/2}}\exp\left(-\frac{(\delta x-u\Delta t)^2}{2 \sigma_\mathrm{diff}^2}\right),
    \label{eq:p_diff_sigma}
\end{equation} 
where $\sigma_\mathrm{diff}$ is a free parameter and can be expressed as a function of the second order moment $\mu_2$ approximated by Eq.\,(\ref{eq:msd})
\begin{equation}
    \begin{split}
        \mu_2 &= \int_{-\infty}^{\infty} P^\mathrm{BDM}(\delta x)(\delta x)^2\,d\delta x\\ 
        &= \int_{-\infty}^{\infty} \exp\left(-\frac{\Delta t}{\tau}\right)P^\mathrm{ball}(\delta x)(\delta x)^2\,d\delta x\\ 
        &+\int_{-\infty}^{\infty}\left[1-\exp\left(-\frac{\Delta t}{\tau}\right)\right]P^\mathrm{diff}(\delta x)(\delta x)^2\,d\delta x\\
        &= \exp\left(-\frac{\Delta t}{\tau}\right)k_BT(\Delta t)^2 \\&+ \left[1-\exp\left(-\frac{\Delta t}{\tau}\right)\right]\sigma_\mathrm{diff}^2.
    \end{split}
\end{equation}
with $\delta x \in X_i$. Now, $\sigma_\mathrm{diff}$ given by
\begin{equation}
    \sigma_\mathrm{diff} = \sqrt{\frac{\mu_2 - \exp\left(-\frac{\Delta t}{\tau}\right)k_BT(\Delta t)^2}{\left[1-\exp\left(-\frac{\Delta t}{\tau}\right)\right]}}.
    \label{eq:sigma_diff}
\end{equation}
We examine the dependence of this diffusion constant on $\Delta t$ in Fig.\,\ref{fig:diffusion}. Our original motivation would demand that $D=\sigma^2_\mathrm{diff}/2\Delta t$ is a constant. However, this is not the case and we will see below that the BDM model only provides a modest improvement over the single Gaussian description. From now on, we will refer to this distribution function as theoretical BDM and denote it as $P^\mathrm{BDM{\text -}T}(X_i)$, since the mean free time $\tau$ and the mean squared displacement are estimated using the velocity auto-correlation function.

\begin{figure}[!htbp]
     \centering
     \subfloat[]{\label{subfig-1:df_bdm}{\includegraphics[width=\columnwidth]{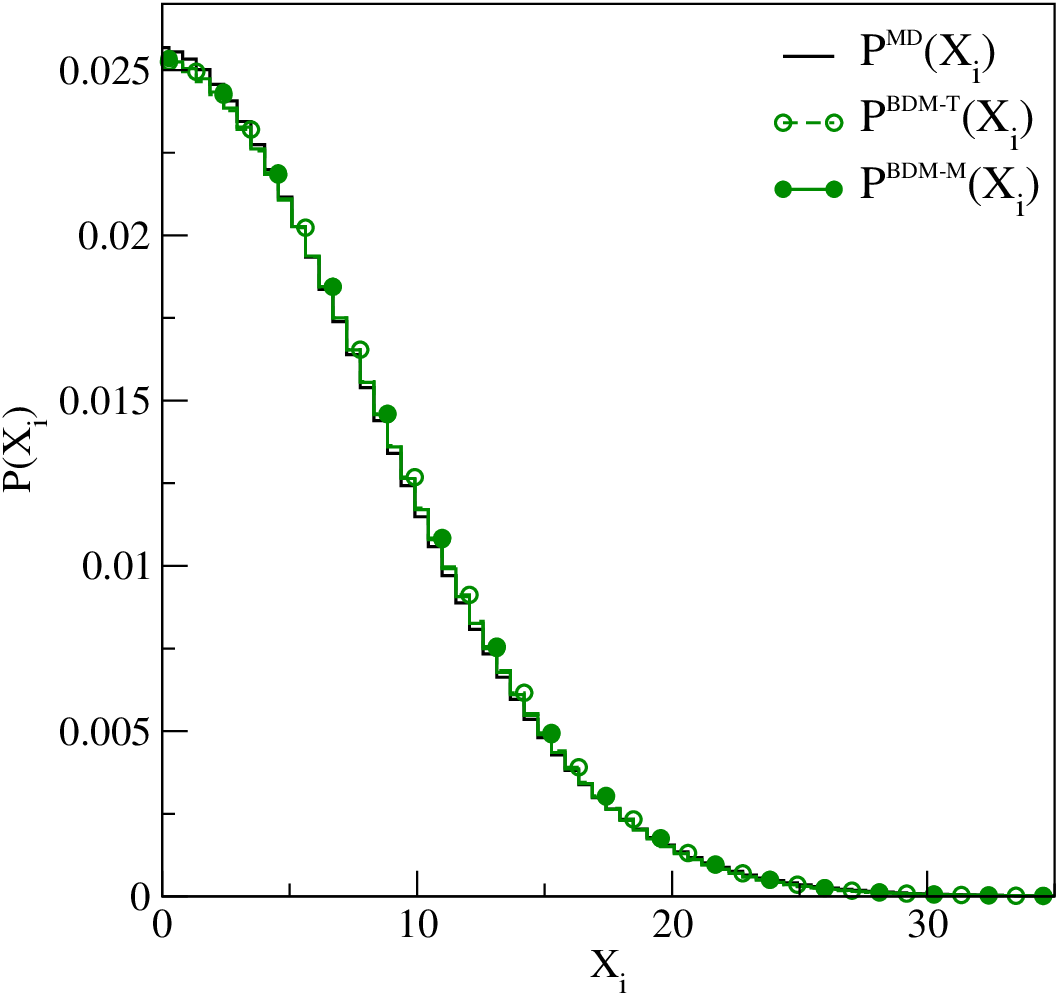}}}\\\hspace{-6mm}
     \subfloat[]{\label{subfig-2:kld_bdm}{\includegraphics[width=\columnwidth]{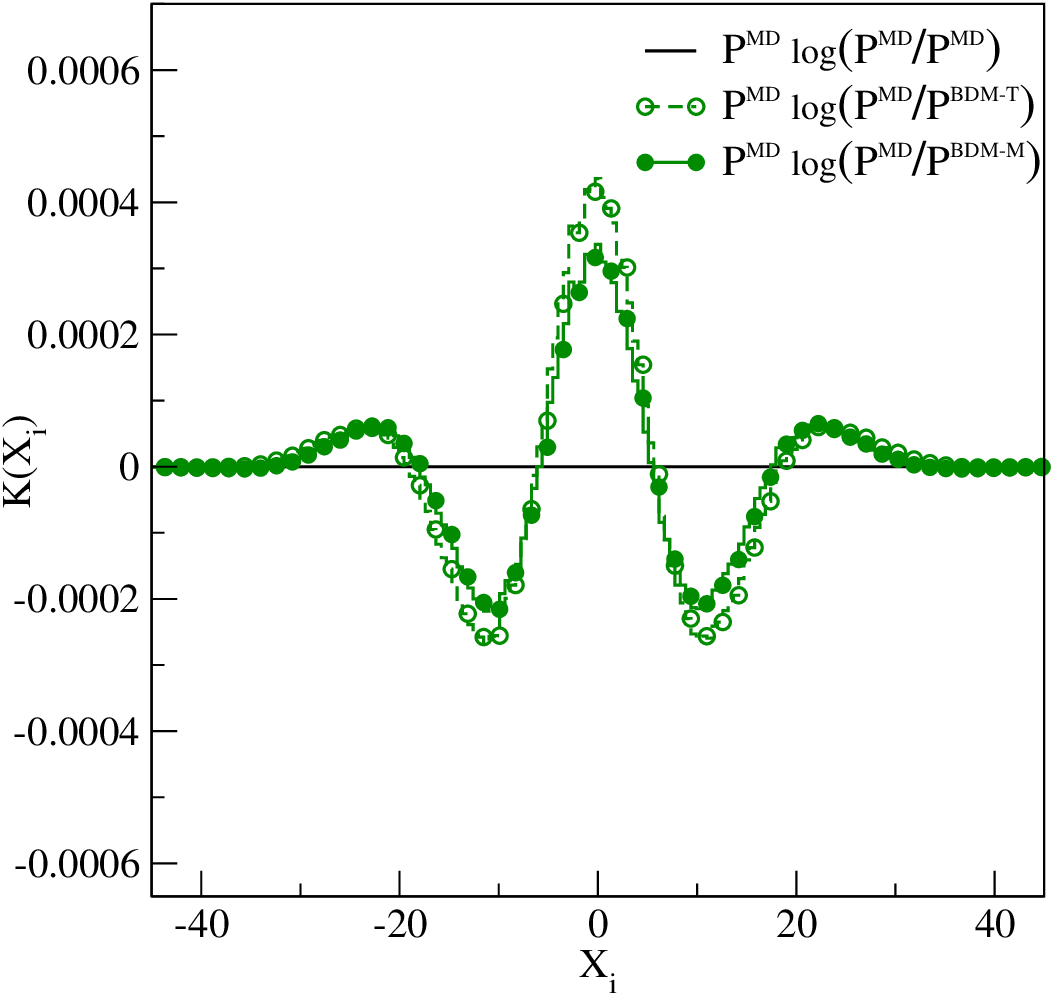}}}
     \caption{(Color online) (a) Displacements probability distribution functions. The solid line (black) depicts a PDF obtained from an MD simulation of LJ particles in equilibrium. The lines with empty or full circles illustrate the ballistic-diffusive distribution function defined in Eq.\,(\ref{eq:p_bdm}) with mean squared displacement obtained from the velocity auto-correlation function as given in Eq.\,(\ref{eq:msd}), and with mean squared displacement fitted directly to the MD data, respectively. Only the data for positive velocities has been depicted due to symmetry. (b) shows the difference between the distributions per interval $X_i$ as defined in Eq.\,(\ref{eq:k_Xi}). The presented data is for the standard parameters used in the paper and a coarse-grained time step $\Delta t = 3.2$. The y-axis has not been re-scaled for a better comparison with Fig.\,\ref{fig:kld_g}.}
     \label{fig:kld_bdm}
\end{figure}

In Fig.\,\ref{subfig-1:df_bdm}, we show the resulting $P^\mathrm{BDM{\text -}T}(X_i)$ distribution function, which resembles well the displacement distribution function obtained from the MD simulation. To assess the discrepancies between the theoretical BDM and the MD distribution function, we calculate $K(P^\mathrm{MD}\parallel P^\mathrm{BDM{\text -}T})$ defined in Eq.\,(\ref{eq:k_Xi}). The results are displayed in Fig.\,\ref{subfig-2:kld_bdm}.
The theoretical ballistic-diffusive probability distribution function with $\Delta t=3.2$ demonstrates a significant improvement in comparison to the single Gaussian distribution function shown in Fig.\,\ref{fig:kld_g}. However, there are noticeable deviations between the PDFs which we will investigate further. 

The second and fourth order moment errors of the BDM distribution function are depicted in Fig.\,\ref{fig:mu_error}. The error is denoted as $\mu^\mathrm{BDM{\text -}T}_2$ (back dashed line) and $\mu^\mathrm{BDM{\text -}T}_4$ (green line with empty circles). The second order moment error is equivalent to $\mu^\mathrm{G{\text -}T}_2$ by construction. This error comes from the long tails of the velocity auto-correlation function shown in Fig.\,\ref{subfig-1:vel_corr}, which are not resolved in the theoretical approximation of the mean squared displacement. Overall, the fourth order moment error of the theoretical BDM model is smaller than the one calculated for the two Gaussian models discussed in Section\,\ref{sec:gaussian}. However, for later times this error increases and becomes as large as the theoretical Gaussian distribution function error.

In order to reduce the error, we construct a second version of the ballistic-diffusive mixture model where we fit the $\mu_2$ and $\mu_4$ moments directly to the MD data. This BDM model does not rely solely on the approximation of the average number of collisions $(\Delta t/\tau)$, which cannot be measured precisely and depends on the approximation made for the velocity auto-correlation function. 

In Sec.\,\ref{sec:gaussian}, we defined the mean squared displacement in terms of the velocity auto-correlation function given by Eq.\,(\ref{eq:def_vcf}). Now, we define the fourth order moment in a similar way
\begin{equation}
\begin{split}
    &\mu_4 = \langle(\delta x_\alpha)^4\rangle\\&=\left\langle\int dt_1v(t_1)\int dt_2v(t_2)\int dt_3v(t_3)\int dt_4v(t_4)\right\rangle\\
    &=\int dt_1\int dt_2\int dt_3\int dt_4\langle v(t_1)v(t_2)v(t_3)v(t_4)\rangle 
\end{split}
\end{equation}
where we need the four-point time correlators for the velocity, that are derived from the displacements given by Eq.\,(\ref{eq:def_displ}). This integral, if feasible, would allow us to calculate theoretically the fourth order moment and thus obtain a better approximation of the probability distribution function of displacements. However, we are unaware of a reliable way to derive this four-point velocity auto-correlation function and therefore, we measure the second and the fourth order moments directly from the MD simulation instead.

We have to  make the following adjustments to the BDM probability distribution function, so that the second and the fourth order moments match the MD data: first, instead of calculating the mean squared displacement from the velocity auto-correlation function, we use the measured mean squared displacement for $\mu_2$ in Eq.\,(\ref{eq:sigma_diff}); second, instead of calculating the mean free path $\tau$ from the velocity auto-correlation function, we define it as a function of $\mu_2$ and $\mu_4$. Thus, the $P^\mathrm{BDM{\text -}M}(X_i)$ distribution function has zero second and fourth order moments error by construction.

The fourth order moment then has the form
\begin{equation}
    \begin{split}
        \mu_4 &= \int_{-\infty}^{\infty} P^\mathrm{BDM}(\delta x)(\delta x)^4\,d\delta x\\ 
              &= \int_{-\infty}^{\infty} \exp\left(-\frac{\Delta t}{\tau}\right)P^\mathrm{ball}(\delta x)(\delta x)^4\,d\delta x\\ 
              &+\int_{-\infty}^{\infty}\left[1-\exp\left(-\frac{\Delta t}{\tau}\right)\right]P^\mathrm{diff}(\delta x)(\delta x)^4\,d\delta x\\ 
              &= 3\left[\sigma_\mathrm{diff}^4 + \exp\left(-\frac{\Delta t}{\tau}\right)\left[(k_BT(\Delta t)^2)^2 - \sigma_\mathrm{diff}^4\right]\right]
    \end{split}
\end{equation}
with $\sigma_\mathrm{diff}$ obtained using the measured second order moment. Now, $\tau$ is not a constant anymore and is given by 
\begin{equation}
    \tau = \ddfrac{-\Delta t}{\ln{\left(\frac{\frac{\mu_4}{3} - \sigma_\mathrm{diff}^4}{[k_BT(\Delta t)^2]^2 - \sigma_\mathrm{diff}^4}\right)}}.
    \label{eq:tau_sigma_diff}
\end{equation}
Eqs.\,(\ref{eq:sigma_diff}) and (\ref{eq:tau_sigma_diff}) define a system of linear equations with two unknowns. The system has a unique solution for $\sigma_\mathrm{diff}$ and $\tau$ as a function of $\mu_2$, $\mu_4$ and $\Delta t$. Thus, a second version of the ballistic-diffusive distribution function is derived and we refer to it as measured ballistic-diffusive distribution function $P^\mathrm{BDM{\text -}M}(X_i)$ because it is fully defined by the MD moments. Details of the derivation are can be found in Appendix\,\ref{app:ball_diff}.

As mentioned earlier, we demand $D=\sigma^2_\mathrm{diff}/2\Delta t$ to be a constant, however, Fig.\,\ref{fig:diffusion} illustrates that $D$ converges to a constant for both BDM models but is not fixed for early time steps. This demonstrates that the BDM model does not capture the physical diffusion properties.

The $P^\mathrm{BDM{\text -}M}(X_i)$ distribution function matches well the MD data as depicted in Fig.\,\ref{subfig-1:df_bdm}. The $K(P^\mathrm{MD}\parallel P^\mathrm{BDM{\text -}T})$ results are illustrated in Fig.\,\ref{subfig-2:kld_bdm} and they show that the divergence between $P^\mathrm{BDM{\text -}M}(X_i)$ and $P^\mathrm{MD}(X_i)$ is smaller in comparison to the theoretical BDM distribution function. However, there is still error with well defined structure, which has to be accounted for.

To gain a better understanding of how the BDM model relates to the MD data and the Gaussian distribution functions, we calculate the KL divergence for $\Delta t \in [0.01,25.6]$ as shown in Fig.\,\ref{fig:kld_error}. The dashed line with empty circles (green) corresponds to the KL divergence $D_\mathrm{KL}(P^\mathrm{MD}\parallel P^\mathrm{BDM{\text -}T})$, while the solid line with full circles (green) illustrates the result of $D_\mathrm{KL}(P^\mathrm{MD}\parallel P^\mathrm{BDM{\text -}M})$. The divergence is decreased by more than half compared to the KL divergence obtained from the Gaussian distribution functions. However, there is still clear error in the intermediate simulation regime.

Even though, we have fitted the second and the fourth order moments, we still have an unsatisfying approximation of the probability distribution function of the displacements. Thus, we conclude that a Gaussian mixture cannot capture the form of the distribution of the displacements for LJ particles in equilibrium. The remaining dependence of $D=\sigma^2_\mathrm{diff}/2\Delta t$ on $\Delta t$ suggests that it is not appropriate to assume that particles that have undergone just one collision will then follow a diffusive displacement. Instead, it might be useful to consider a range of distribution functions occurring after a number of collisions. We will follow up this idea in the next section.

\section{Poisson Weighted Sum of Gaussian Distribution Functions}
\label{sec:poisson_gaussians}
The number of collisions within a time interval plays an important role in the definition of the probability distribution function of displacements. We can prove this statement by a thought experiment: consider a number of particles in a domain. When the particles undergo a collision their direction and velocity changes. This in turn means that the collisions also change the probability of certain displacements to occur. 

In this section, we assume that the intermediate ballistic-diffusive regime could be described as a Poisson weighted sum of Gaussian distributions. One can consider that after a time step $\Delta t$ the particles can be divided into groups depending on the number of collisions they have experienced. We model these particle collisions using the Poisson probability distribution function
\begin{equation}
    P(\delta x) = \sum_{c=0}^\infty e^{-\lambda} \frac{\lambda^c}{c!},
\end{equation}
$\lambda$ is effectively the average number of collisions given by 
\begin{equation}
    \lambda=\frac{\Delta t}{\tau}
    \label{eq:lambda_def}
\end{equation}
where $\tau\approx 0.728$ is the mean free time obtained using Eq.\,(\ref{eq:vel_corr_exp_decay}). In this formulation the mean free time is considered to be an exponential decay constant. In principle the timing of the collisions should also be random (i.e. given by a Poisson process), but the resulting integrals over the collision times do not admit analytical solutions. Assuming that the collisions are evenly spaced may introduce a small error, but it makes the resulting displacements after $c$ collisions again Gaussian, which simplifies the application of our results. For details on arbitrary collision occurring at random time refer to Appendix\,\ref{app:poisson}.

With this approximation the Poisson Weighted Sum of Gaussians (WSG) model is then given as 
\begin{equation}
    \begin{split}
    P^{\mathrm{WSG}}(\delta x) &= \sum_{c=0}^\infty e^{-\lambda} \frac{\lambda^c}{c!} \frac{\sqrt{(\lambda+1)}}{\sqrt{2\pi (c+1)\langle(\delta x)^2\rangle}}\\ 
    &\times\exp\left(-\frac{(\lambda+1)(\delta x-u\Delta t)^2}{2(c+1)\langle (\delta x)^2\rangle}\right)
    \end{split}
    \label{eq:poisson_v2}
\end{equation}
for displacements $\delta x$ in one dimension.
In extreme regimes, being purely ballistic or purely diffusive, the probability distribution function $P^\mathrm{WSG}(\delta x)$ is reduced to a single Gaussian distribution given by Eq.\,(\ref{eq:p_ball}) or Eq.\,(\ref{eq:p_diff}), respectively. However, in an intermediate regime, we will have contributions from multiple Gaussian distribution functions weighted by a Poisson distribution function. 

\begin{figure*}[!htbp]
     \centering
     \subfloat[\label{subfig-1:df_wsg}]{{\includegraphics[width=\columnwidth]{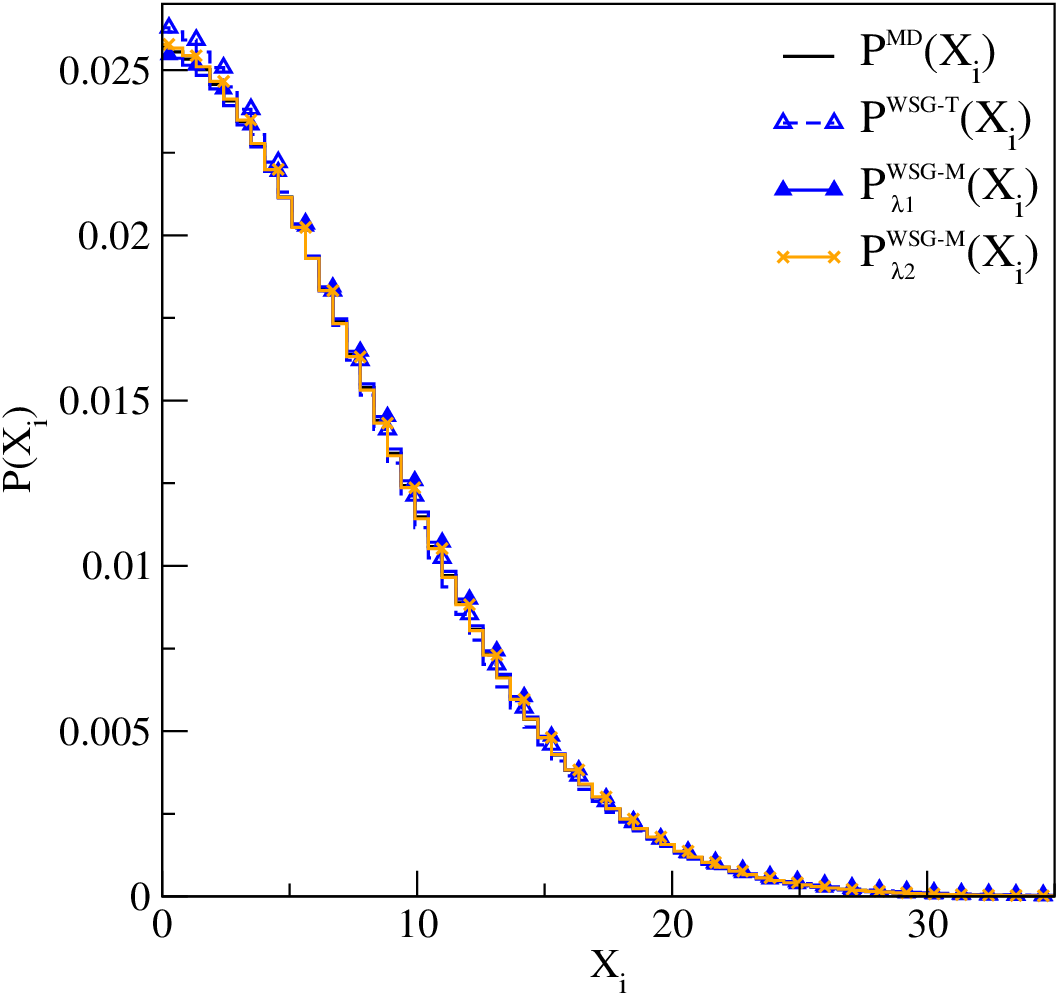}}}\hspace{1em}
     \subfloat[\label{subfig-2:kld_0.01_wsg}]{{\includegraphics[width=\columnwidth]{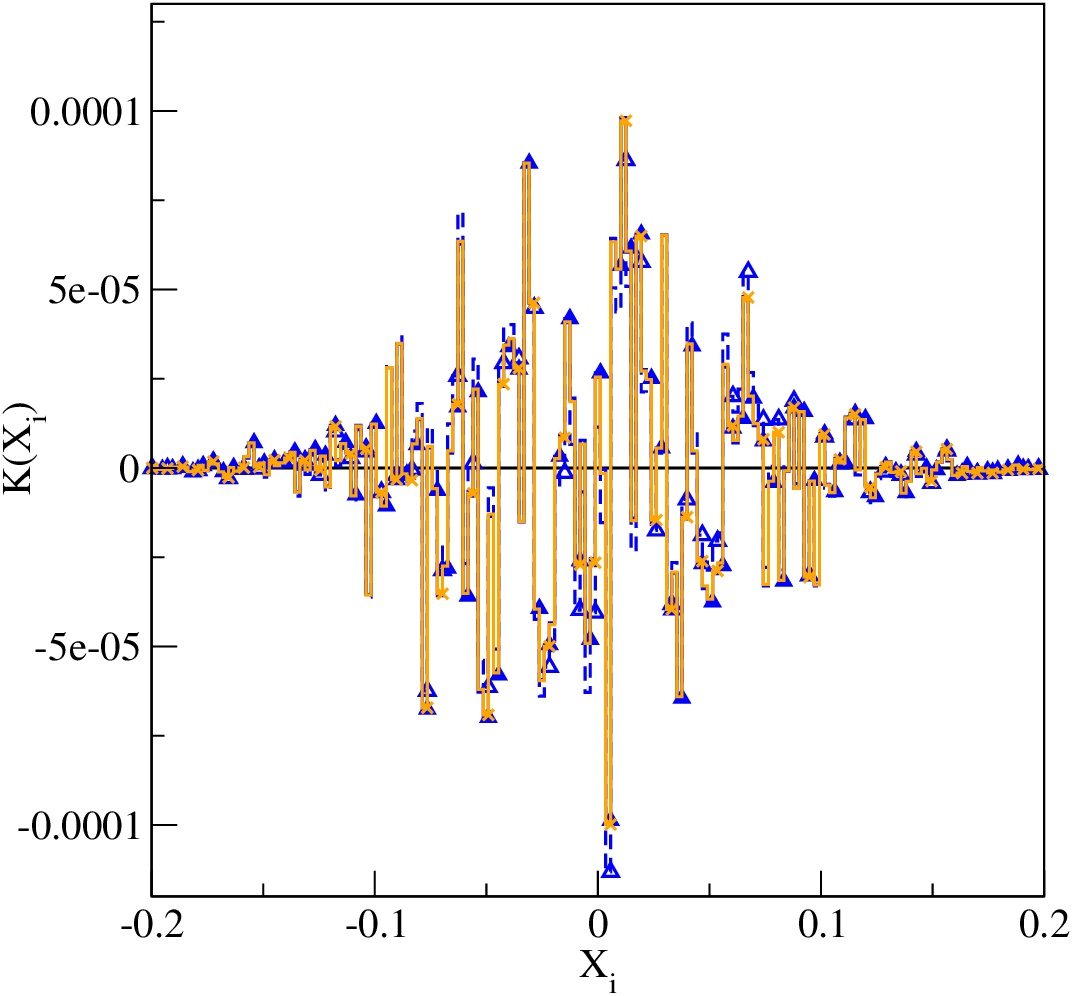}}}\\
     \subfloat[\label{subfig-2:kld_0.1_wsg}]{{\includegraphics[width=\columnwidth]{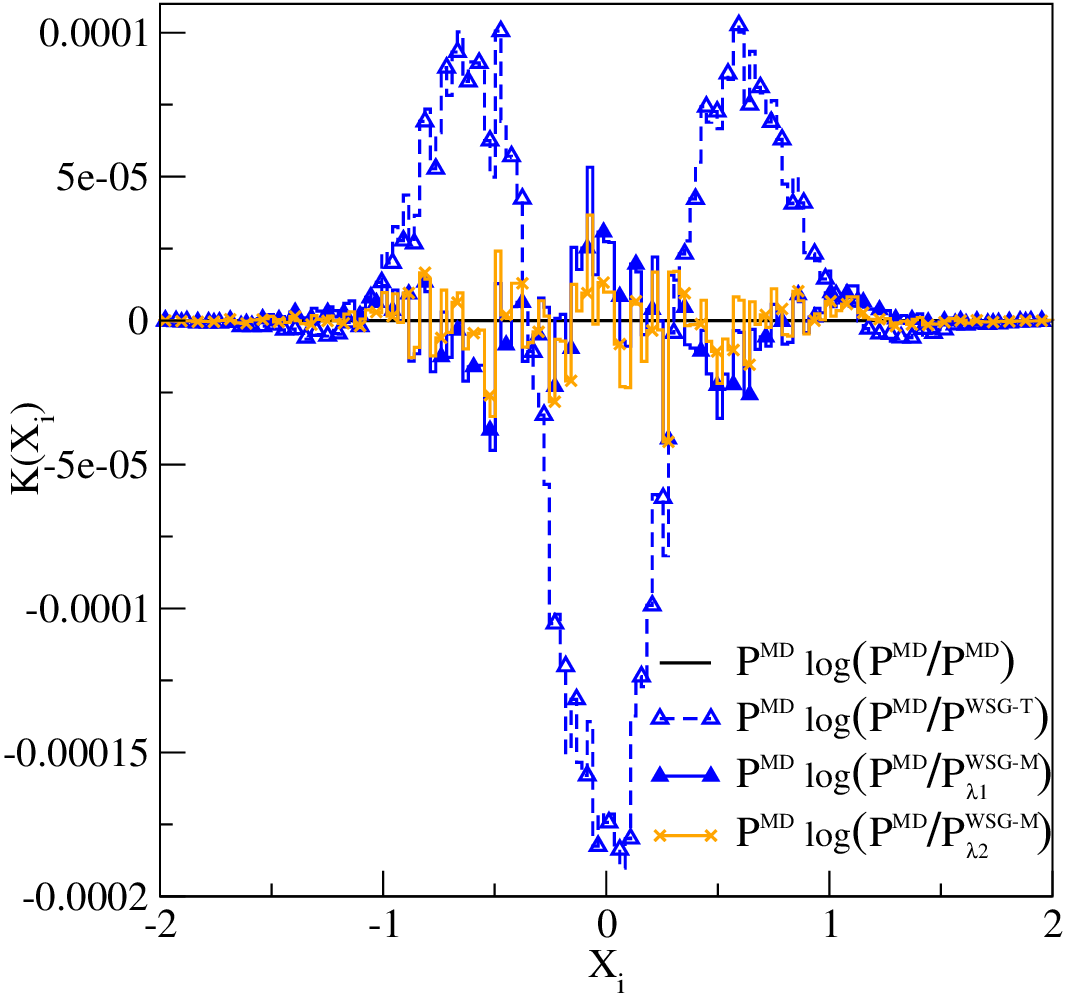}}}\hspace{1em}
     \subfloat[\label{subfig-2:kld_wsg}]{{\includegraphics[width=\columnwidth]{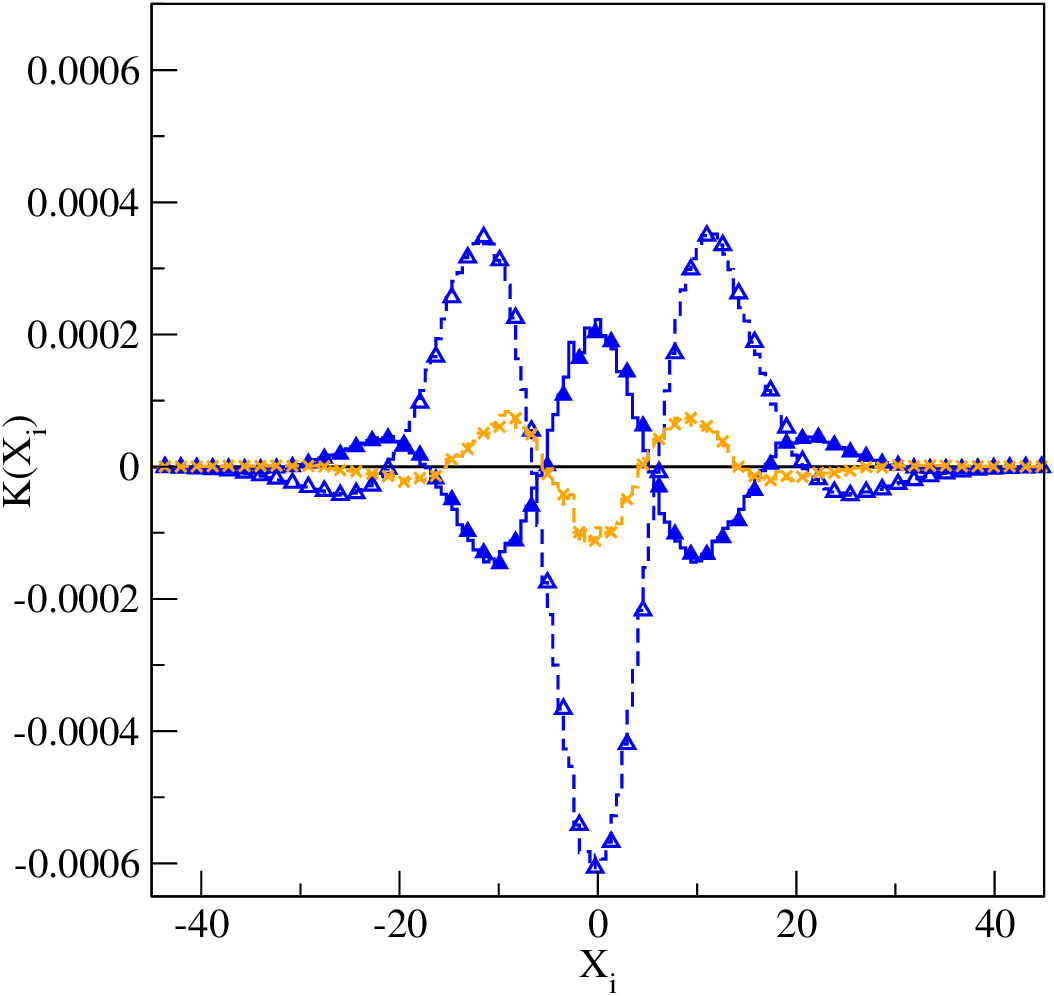}}}
     \caption{(Color online) (a) Displacements probability distribution functions. The solid line (black) depicts a PDF obtained from an MD simulation of LJ particles in equilibrium. The dashed line (blue) with empty triangles illustrates the PDF of the theoretical WSG defined in Eq.\,(\ref{eq:poisson_v2}) with $\lambda$ obtained using the theoretical velocity auto-correlation function from Eq.\,(\ref{eq:vel_corr_exp_decay}). The solid lines with full squares or x-symbols denote the Poisson WSG distribution function with average number of collisions $\lambda_1$ and $\lambda_2$, respectively. The time step is $\Delta t = 3.2$ and due to symmetry only the data for positive velocities has been depicted. (b)-(d) show the difference between the distributions per interval $X_i$ as defined in Eq.\,(\ref{eq:k_Xi}) for a variety of time steps: (b)\,$\Delta t = 0.01$, (c)\,$\Delta t = 0.1$, and (d)\,$\Delta t = 3.2$. The presented data is for the standard parameters used in the paper. The y-axis of (a) and (d) have not been re-scaled for a better comparison with Fig.\,\ref{fig:kld_g} and Fig.\,\ref{fig:kld_bdm}.}
     \label{fig:kld_wsg}
\end{figure*}

By using the definition of the average number of collisions given in Eq.\,(\ref{eq:lambda_def}) and obtaining the mean free time and the mean squared displacement based on the velocity auto-correlation function, we recover a fully defined theoretical version of the Poisson WSG model which we refer to as $P^{\mathrm{WSG\text -T}}(X_i)$. This probability distribution function is illustrated in Fig.\,\ref{subfig-1:df_wsg} by a dashed line with empty triangles (blue). $P^{\mathrm{WSG\text -T}}(X_i)$ shows a good fit to the distribution function measured directly from the MD simulation but there are still visible discrepancies.

In Figs.\,\ref{subfig-2:kld_0.01_wsg}-\ref{subfig-2:kld_wsg}, the $K(P^\mathrm{MD}\parallel P^\mathrm{WSG{\text -}T})$ function is illustrated for three different time steps: $\Delta t=0.01$, $\Delta t=0.1$ and $\Delta t=3.2$. The results for $\Delta t=0.01$ show noise coming from the averaging procedure as one can see in Fig.\,\ref{subfig-2:kld_0.01_wsg}. With increasing the time step, we start seeing some structure in the discrepancies between the theoretical weighted sum of Gaussians and the MD probability distribution function as shown in Fig.\,\ref{subfig-2:kld_0.1_wsg}. For $\Delta t=3.2$, one can see that the rate of discrepancies is as large as the one shown in Fig.\,\ref{subfig-2:kld_g} calculated for the single Gaussian distribution function but with an opposite sign. This observation suggests that the theoretical WSG does not capture well the distribution of the measured MD displacements. 

To compare the overall performance of $P^\mathrm{WSG{\text -}T}(X)$, we calculate its KL divergence as shown in Fig.\,\ref{fig:kld_error} (blue dashed line with empty triangles). The $D_\mathrm{KL}(P^\mathrm{MD}\parallel P^\mathrm{WSG{\text -}T})$ divergence is slightly smaller than the one measured for the Gaussian models presented in Sec.\,\ref{sec:gaussian}.  

In order to find the source of the large KL divergence, we display the second and fourth order moments error in Fig.\,\ref{fig:mu_error}. The second order moment error is equivalent to the error calculated for the other theoretical models ($\mu_2^\mathrm{G{\text -}T}\text{= }\mu_2^\mathrm{BDM{\text -}T}\text{= }\mu_2^\mathrm{WSG{\text -}T}$). The $P^\mathrm{WSG{\text-}T}(X_i)$ fourth order moment error, however, is larger than the fourth order moment error of the other two models. This is true especially for the intermediate regime and explains the poor results of the theoretical WSG model. 

The average number of collisions $\lambda$ plays an important role in the definition of the BDM and WSG models. Unfortunately, it is difficult to make a good approximation for $\lambda$ based on the velocity auto-correlation function. Therefore, to reduce the error coming from the theoretical average number of collisions and to eliminate the second and the fourth order moment errors, we match these moments to the corresponding moments measured directly from the MD simulations. We derive the mean squared displacement from the second order Gaussian integral
\begin{equation}
    \begin{split}
        \mu_2 &=\int_{-\infty}^\infty P^\mathrm{WSG}(\delta x)(\delta x)^2\, d\delta x\\ 
        &=  \int_{-\infty}^\infty \sum_{c=0}^\infty e^{-\lambda}\frac{\lambda^c}{c!} \frac{\sqrt{\lambda+1}}{\sqrt{2\pi(c+1)\langle(\delta x)^2\rangle}} \\ 
        &\times\exp\left(-\frac{(\lambda+1)(\delta x-u\Delta t)^2}{2(c+1)\langle(\delta x)^2\rangle}\right) (\delta x)^2\,d\delta x
    \end{split}
\end{equation}
and the fourth order moment from the fourth order Gaussian integral
\begin{equation}
    \begin{split}
    \mu_4 &= \int_{-\infty}^\infty P^\mathrm{WSG}(\delta x)(\delta x)^4 \,d(\delta x)\\ 
    &= \int_{-\infty}^\infty \sum_{c=0}^\infty e^{-\lambda} \frac{\lambda^c}{c!} \frac{\sqrt{\lambda+1}}{\sqrt{2\pi (c+1)\langle(\delta x)^2\rangle}} \\ 
    &\times\exp\left(-\frac{(\lambda+1)(\delta x-u\Delta t)^2}{2(c+1)\langle(\delta x)^2\rangle}\right) (\delta x)^4 \,d\delta x\\ 
    &=\frac{3 \langle(\delta x)^2\rangle^2}{(\lambda+1)^2}\left[\lambda^2+3\lambda+1\right].
   \end{split}
   \label{eq:fourth_order}
\end{equation}
This ensures that the $\mu_2$ and $\mu_4$ moments are fully recovered from the WSG model. Now, we can express $\lambda$ as a function of these parameters and solve the resulting quadratic equation
\begin{equation}
    \frac{3 \mu_2^2}{(\lambda+1)^2}\left[\lambda^2+3\lambda+1\right] -\mu_4 = 0
    \label{eq:lambda_eq}
\end{equation}
with $\mu_2=\langle(\delta x)^2\rangle$ for brevity. The quadratic equation has the following solutions 
\begin{equation}
   \hspace{-3mm}\lambda_{1,2} =\frac{-9\mu_2^2\pm\sqrt{3[15\mu_2^4-4\mu_2^2\mu_4]}+2\mu_4}{2[3\mu_2^2-\mu_4]}.
\end{equation}
Details of the derivation are omitted but they can be found in Appendix \ref{app:poisson}.

\begin{figure}[!t]
\includegraphics[width=\columnwidth]{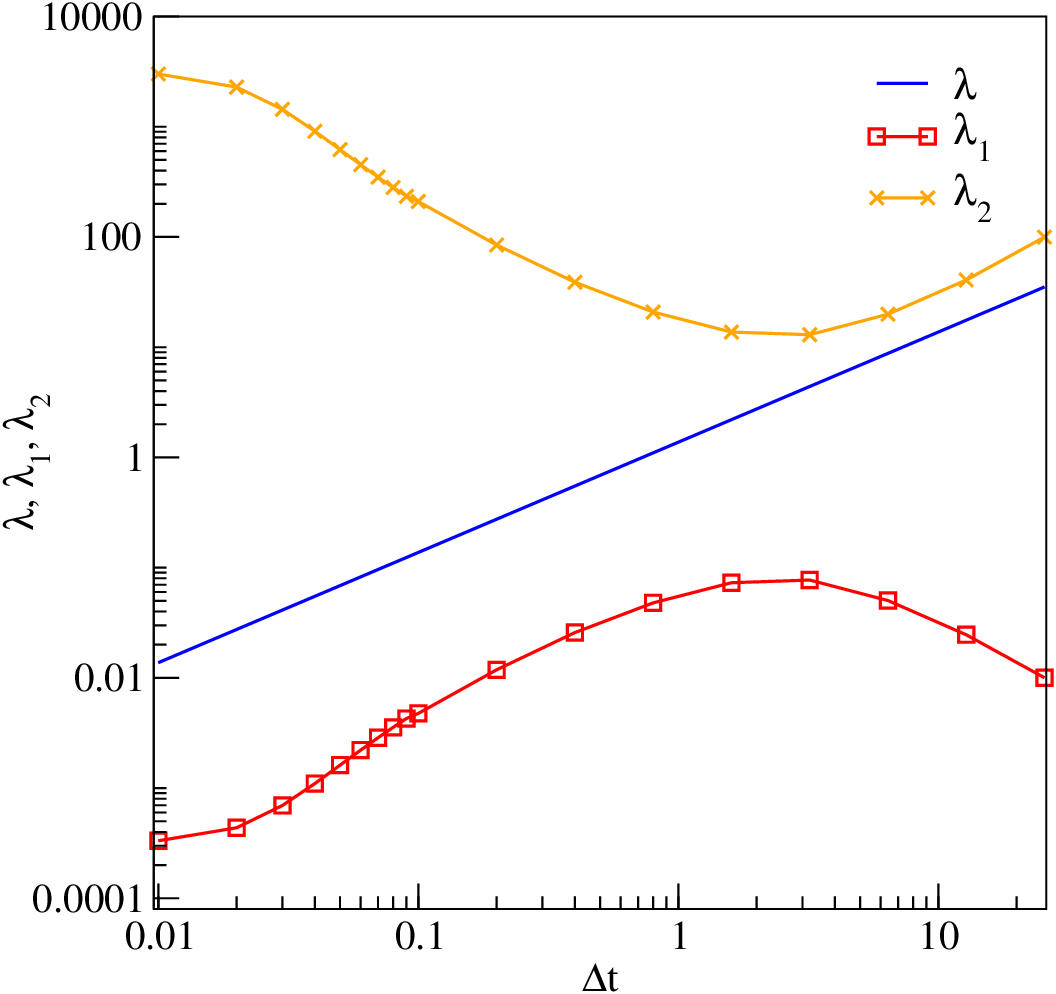}%
\caption{(Color online) Average number of collisions depending on the coarse-grained time step $\Delta t$. $\lambda$ denotes the number of collisions obtained from the velocity auto-correlation theory given in Eq.\,(\ref{eq:lambda_def}), which is used for the calculation of the $P^\mathrm{WSG{\text -}T}(X_i)$ distribution function. $\lambda_1$ and $\lambda_2$ are solutions of the quadratic equation given in Eq.\,(\ref{eq:lambda_eq}). These values are used for the calculation of $P_{\lambda_1}^\mathrm{WSG{\text -}M}(X_i)$ and $P_{\lambda_2}^\mathrm{WSG{\text -}M}(X_i)$ distribution functions, respectively. $\lambda_1$ and $\lambda_2$ are obtained using the second and the fourth order moments measured directly from the MD simulations.}
\label{fig:lambda}
\end{figure}

Since the mean squared displacement and the fourth order moment depend wholly on the time step, we plot $\lambda_1(\Delta t)$, $\lambda_2(\Delta t)$, and $\lambda(\Delta t)$ from Eq.\,(\ref{eq:lambda_def}) as a function of $\Delta t$, which is depicted in Fig.\,\ref{fig:lambda}. 
We see that the analytical expectation for $\lambda$ from Eq.\,(\ref{eq:lambda_def}) is in better agreement with $\lambda_2$ for large $\Delta t$, while for small $\Delta t$ it is in better agreement with $\lambda_1$. This is intriguing, and we do not fully understand the significance of this result. However, we should note here that both limits $\Delta t\rightarrow 0$ and $\Delta t\rightarrow \infty$ lead to a simple Gaussian distribution. For small $\Delta t$ this is the case because there is only the $c=0$ term in the Poisson distribution matters, and for large $\Delta t$ because the Poisson distribution will be sharply peaked around $c=\lambda$, leading again to a simple Gaussian distribution function.

$P^\mathrm{WSG{\text -}M}(X_i)$ has zero second and fourth order moment errors by construction, because these moments have been fitted to the MD simulation data.

The Kullback-Leibler divergence per element $X_i$ for $P_{\lambda_1}^\mathrm{WSG{\text -}M}(X_i)$ and $P_{\lambda_2}^\mathrm{WSG{\text -}M}(X_i)$ is illustrated in Figs.\,\ref{subfig-2:kld_0.01_wsg}-\ref{subfig-2:kld_wsg}. In each figure, $K(P^\mathrm{MD}\parallel P_{\lambda_1}^\mathrm{WSG{\text -}M})$ and $K(P^\mathrm{MD}\parallel P_{\lambda_2}^\mathrm{WSG{\text -}M})$ are depicted for different time step. Fig.\,\ref{subfig-2:kld_0.01_wsg} shows the error of $K(P^\mathrm{MD}\parallel P_{\lambda_2}^\mathrm{WSG{\text -}M})$ for $\Delta t=0.01$ where the error is very small and is dominated by noise due to the averaging procedure. For the coarse-grained time step of $\Delta t=0.1$, the error becomes larger and one sees small structures building, however, the noise is still dominant in the error contribution. In Fig.\,\ref{subfig-2:kld_wsg}, we show the $K(P^\mathrm{MD}\parallel P_{\lambda_2}^\mathrm{WSG{\text -}M})$ results for $\Delta t=3.2$. There is a clear structure of the error for both WSG-M probability distribution functions. In comparison to the Gaussian and the Ballistic-Diffusive mixture models, the WSG model with average number of collisions $\lambda_1$ and $\lambda_2$ shows much smaller error.

For better comparison, we calculate the Kullback-Leibler divergence for $P_{\lambda_1}^\mathrm{WSG{\text -}M}(X_i)$ and $P_{\lambda_2}^\mathrm{WSG{\text -}M}(X_i)$ and display the results in Fig.\,\ref{fig:kld_error}. The KL divergence for $\lambda_1$ shows reduced error for the transition regime. The second solution of Eq.\,(\ref{eq:lambda_eq}) $\lambda_2$, however, shows KL divergence close to zero for all time steps. This is a significant improvement comparing the results using a single Gaussian or a mixture of two Gaussian distribution functions.

The WSG probability distribution function strongly depends on the calculation of the average number of collisions. By using the theoretical average number of collisions obtained from the velocity auto-correlation function, the KL divergence $D_\mathrm{KL}(P^\mathrm{MD}\parallel P^\mathrm{WSG{\text -}T})$ is almost as large as $D_\mathrm{KL}(P^\mathrm{MD}\parallel P^\mathrm{G})$ for the Gaussian models. Even fitting the second and the fourth order moments is not sufficient to obtain a good estimation of the PDF obtained from the MD simulation. The KL divergence for the WSG model with $\lambda_1$ shows an improvement by about a factor of 6 but it still large in the transition regime. $P^\mathrm{WSG{\text -}M}(X_i)$ with $\lambda_2$ gives a unique close to zero Kullback-Leibler divergence owing to the WSG model and the correct choice of the average number of collisions.
\section{Conclusions}
\label{sec:conclusions}
In this article we have shown that displacement distributions are only of a Gaussian form for either very small times or for long times. The transition region, where a different distribution function is found roughly corresponds to the region where the motion of particles transitions from ballistic to diffusive regime. One signal of the deviation is the fourth order moment of the probability distribution function of displacements. 

By allowing for the distribution to be a mixture of two distribution functions, one corresponding to the ballistic regime, and a second one to be selected to give the correct second and fourth order moments gives a PDF that agrees better with the MD distribution function, by about a factor of 3 measured by the Kullback-Leibler divergence. 

Using the same amount of information, i.e. the second and fourth order moments, we found a different distribution function that gives a nearly perfect fit. This distribution was motivated by considering the distribution function as a mixture of Gaussian distributions that have undergone a number of collisions, which are given by a Poisson distribution. 

This analytical description is very promising for the MDLG analysis of collision operators in non-equilibrium systems. It would be very helpful if a theoretical prediction of the fourth order moment equivalent to the second order moment derived from the velocity time correlation could be achieved, because then one could obtain the displacement distribution for all time steps through one measurement. The current approach still needs measurements of the fourth order moment for each time step. Furthermore, the current study was done for a semi-dilute system. In future research, we anticipate to establish up to what density the distribution with Poisson weighted sum of Gaussians remains a valid description for the displacement distribution.

\begin{acknowledgments}
AW and AP would like to thank M. Reza Parsa and Michael E. Cates for the helpful discussions. AP was partially supported by the Center for Nonlinear Studies (CNLS) and the Laboratory Directed Research and Development (LDRD) program at Los Alamos National Laboratory (LANL), and the German Federal Ministry of Education and Research (BMBF) in the scope of the project Aerotherm (reference numbers: 01IS16016A-B).
\end{acknowledgments}


\bibliography{references_05_06}

\begin{thebibliography}{20}%
\makeatletter
\providecommand \@ifxundefined [1]{%
 \@ifx{#1\undefined}
}%
\providecommand \@ifnum [1]{%
 \ifnum #1\expandafter \@firstoftwo
 \else \expandafter \@secondoftwo
 \fi
}%
\providecommand \@ifx [1]{%
 \ifx #1\expandafter \@firstoftwo
 \else \expandafter \@secondoftwo
 \fi
}%
\providecommand \natexlab [1]{#1}%
\providecommand \enquote  [1]{``#1''}%
\providecommand \bibnamefont  [1]{#1}%
\providecommand \bibfnamefont [1]{#1}%
\providecommand \citenamefont [1]{#1}%
\providecommand \href@noop [0]{\@secondoftwo}%
\providecommand \href [0]{\begingroup \@sanitize@url \@href}%
\providecommand \@href[1]{\@@startlink{#1}\@@href}%
\providecommand \@@href[1]{\endgroup#1\@@endlink}%
\providecommand \@sanitize@url [0]{\catcode `\\12\catcode `\$12\catcode
  `\&12\catcode `\#12\catcode `\^12\catcode `\_12\catcode `\%12\relax}%
\providecommand \@@startlink[1]{}%
\providecommand \@@endlink[0]{}%
\providecommand \url  [0]{\begingroup\@sanitize@url \@url }%
\providecommand \@url [1]{\endgroup\@href {#1}{\urlprefix }}%
\providecommand \urlprefix  [0]{URL }%
\providecommand \Eprint [0]{\href }%
\providecommand \doibase [0]{https://doi.org/}%
\providecommand \selectlanguage [0]{\@gobble}%
\providecommand \bibinfo  [0]{\@secondoftwo}%
\providecommand \bibfield  [0]{\@secondoftwo}%
\providecommand \translation [1]{[#1]}%
\providecommand \BibitemOpen [0]{}%
\providecommand \bibitemStop [0]{}%
\providecommand \bibitemNoStop [0]{.\EOS\space}%
\providecommand \EOS [0]{\spacefactor3000\relax}%
\providecommand \BibitemShut  [1]{\csname bibitem#1\endcsname}%
\let\auto@bib@innerbib\@empty
\bibitem [{\citenamefont {Ermak}\ and\ \citenamefont
  {McCammon}(1978)}]{ermak1978brownian}%
  \BibitemOpen
  \bibfield  {author} {\bibinfo {author} {\bibfnamefont {D.~L.}\ \bibnamefont
  {Ermak}}\ and\ \bibinfo {author} {\bibfnamefont {J.~A.}\ \bibnamefont
  {McCammon}},\ }\href@noop {} {\bibfield  {journal} {\bibinfo  {journal} {The
  Journal of chemical physics}\ }\textbf {\bibinfo {volume} {69}},\ \bibinfo
  {pages} {1352} (\bibinfo {year} {1978})}\BibitemShut {NoStop}%
\bibitem [{\citenamefont {Hoogerbrugge}\ and\ \citenamefont
  {Koelman}(1992)}]{hoogerbrugge1992simulating}%
  \BibitemOpen
  \bibfield  {author} {\bibinfo {author} {\bibfnamefont {P.}~\bibnamefont
  {Hoogerbrugge}}\ and\ \bibinfo {author} {\bibfnamefont {J.}~\bibnamefont
  {Koelman}},\ }\href@noop {} {\bibfield  {journal} {\bibinfo  {journal} {EPL
  (Europhysics Letters)}\ }\textbf {\bibinfo {volume} {19}},\ \bibinfo {pages}
  {155} (\bibinfo {year} {1992})}\BibitemShut {NoStop}%
\bibitem [{\citenamefont {Ihle}\ and\ \citenamefont
  {Kroll}(2001)}]{ihle2001stochastic}%
  \BibitemOpen
  \bibfield  {author} {\bibinfo {author} {\bibfnamefont {T.}~\bibnamefont
  {Ihle}}\ and\ \bibinfo {author} {\bibfnamefont {D.}~\bibnamefont {Kroll}},\
  }\href@noop {} {\bibfield  {journal} {\bibinfo  {journal} {Physical Review
  E}\ }\textbf {\bibinfo {volume} {63}},\ \bibinfo {pages} {020201} (\bibinfo
  {year} {2001})}\BibitemShut {NoStop}%
\bibitem [{\citenamefont {Qian}\ \emph {et~al.}(1992)\citenamefont {Qian},
  \citenamefont {D'Humières},\ and\ \citenamefont
  {Lallemand}}]{qian_lattice_1992}%
  \BibitemOpen
  \bibfield  {author} {\bibinfo {author} {\bibfnamefont {Y.~H.}\ \bibnamefont
  {Qian}}, \bibinfo {author} {\bibfnamefont {D.}~\bibnamefont {D'Humières}},\
  and\ \bibinfo {author} {\bibfnamefont {P.}~\bibnamefont {Lallemand}},\ }\href
  {https://doi.org/10.1209/0295-5075/17/6/001} {\bibfield  {journal} {\bibinfo
  {journal} {Europhysics Letters (EPL)}\ }\textbf {\bibinfo {volume} {17}},\
  \bibinfo {pages} {479} (\bibinfo {year} {1992})}\BibitemShut {NoStop}%
\bibitem [{\citenamefont {He}\ and\ \citenamefont {Luo}(1997)}]{he1997theory}%
  \BibitemOpen
  \bibfield  {author} {\bibinfo {author} {\bibfnamefont {X.}~\bibnamefont
  {He}}\ and\ \bibinfo {author} {\bibfnamefont {L.-S.}\ \bibnamefont {Luo}},\
  }\href@noop {} {\bibfield  {journal} {\bibinfo  {journal} {Physical Review
  E}\ }\textbf {\bibinfo {volume} {56}},\ \bibinfo {pages} {6811} (\bibinfo
  {year} {1997})}\BibitemShut {NoStop}%
\bibitem [{\citenamefont {Parsa}\ and\ \citenamefont
  {Wagner}(2017)}]{parsa_lattice_2017}%
  \BibitemOpen
  \bibfield  {author} {\bibinfo {author} {\bibfnamefont {M.~R.}\ \bibnamefont
  {Parsa}}\ and\ \bibinfo {author} {\bibfnamefont {A.~J.}\ \bibnamefont
  {Wagner}},\ }\href {https://doi.org/10.1103/PhysRevE.96.013314} {\bibfield
  {journal} {\bibinfo  {journal} {Physical Review E}\ }\textbf {\bibinfo
  {volume} {96}},\ \bibinfo {pages} {013314} (\bibinfo {year}
  {2017})}\BibitemShut {NoStop}%
\bibitem [{\citenamefont {Parsa}\ \emph {et~al.}(2019)\citenamefont {Parsa},
  \citenamefont {Pachalieva},\ and\ \citenamefont
  {Wagner}}]{parsa_validity_2019}%
  \BibitemOpen
  \bibfield  {author} {\bibinfo {author} {\bibfnamefont {M.~R.}\ \bibnamefont
  {Parsa}}, \bibinfo {author} {\bibfnamefont {A.}~\bibnamefont {Pachalieva}},\
  and\ \bibinfo {author} {\bibfnamefont {A.~J.}\ \bibnamefont {Wagner}},\
  }\href {https://doi.org/10.1142/S0129183119410079} {\bibfield  {journal}
  {\bibinfo  {journal} {International Journal of Modern Physics C}\ }\textbf
  {\bibinfo {volume} {30}},\ \bibinfo {pages} {1941007} (\bibinfo {year}
  {2019})}\BibitemShut {NoStop}%
\bibitem [{\citenamefont {Parsa}\ and\ \citenamefont
  {Wagner}(2020)}]{parsa2020}%
  \BibitemOpen
  \bibfield  {author} {\bibinfo {author} {\bibfnamefont {M.~R.}\ \bibnamefont
  {Parsa}}\ and\ \bibinfo {author} {\bibfnamefont {A.~J.}\ \bibnamefont
  {Wagner}},\ }\href {https://doi.org/10.1103/PhysRevLett.124.234501}
  {\bibfield  {journal} {\bibinfo  {journal} {Phys. Rev. Lett.}\ }\textbf
  {\bibinfo {volume} {124}},\ \bibinfo {pages} {234501} (\bibinfo {year}
  {2020})}\BibitemShut {NoStop}%
\bibitem [{\citenamefont {Masoliver}(2017)}]{masoliver_three-dimensional_2017}%
  \BibitemOpen
  \bibfield  {author} {\bibinfo {author} {\bibfnamefont {J.}~\bibnamefont
  {Masoliver}},\ }\href {https://doi.org/10.1103/PhysRevE.96.022101} {\bibfield
   {journal} {\bibinfo  {journal} {Physical Review E}\ }\textbf {\bibinfo
  {volume} {96}},\ \bibinfo {pages} {022101} (\bibinfo {year}
  {2017})}\BibitemShut {NoStop}%
\bibitem [{\citenamefont {Masoliver}\ and\ \citenamefont
  {Lindenberg}(2020)}]{masoliver_two-dimensional_2020}%
  \BibitemOpen
  \bibfield  {author} {\bibinfo {author} {\bibfnamefont {J.}~\bibnamefont
  {Masoliver}}\ and\ \bibinfo {author} {\bibfnamefont {K.}~\bibnamefont
  {Lindenberg}},\ }\href {https://doi.org/10.1103/PhysRevE.101.012137}
  {\bibfield  {journal} {\bibinfo  {journal} {Physical Review E}\ }\textbf
  {\bibinfo {volume} {101}},\ \bibinfo {pages} {012137} (\bibinfo {year}
  {2020})}\BibitemShut {NoStop}%
\bibitem [{\citenamefont {Bhatnagar}\ \emph {et~al.}(1954)\citenamefont
  {Bhatnagar}, \citenamefont {Gross},\ and\ \citenamefont
  {Krook}}]{bhatnagar_model_1954}%
  \BibitemOpen
  \bibfield  {author} {\bibinfo {author} {\bibfnamefont {P.~L.}\ \bibnamefont
  {Bhatnagar}}, \bibinfo {author} {\bibfnamefont {E.~P.}\ \bibnamefont
  {Gross}},\ and\ \bibinfo {author} {\bibfnamefont {M.}~\bibnamefont {Krook}},\
  }\href {https://doi.org/10.1103/PhysRev.94.511} {\bibfield  {journal}
  {\bibinfo  {journal} {Physical Review}\ }\textbf {\bibinfo {volume} {94}},\
  \bibinfo {pages} {511} (\bibinfo {year} {1954})}\BibitemShut {NoStop}%
\bibitem [{\citenamefont {Pachalieva}\ and\ \citenamefont
  {Wagner}(shed)}]{pachalieva_unpublished}%
  \BibitemOpen
  \bibfield  {author} {\bibinfo {author} {\bibfnamefont {A.}~\bibnamefont
  {Pachalieva}}\ and\ \bibinfo {author} {\bibfnamefont {A.}~\bibnamefont
  {Wagner}},\ }\href@noop {} {\  (\bibinfo {year} {unpublished})}\BibitemShut
  {NoStop}%
\bibitem [{\citenamefont {Plimpton}(1995)}]{plimpton_fast_1995}%
  \BibitemOpen
  \bibfield  {author} {\bibinfo {author} {\bibfnamefont {S.}~\bibnamefont
  {Plimpton}},\ }\href {https://doi.org/10.1006/jcph.1995.1039} {\bibfield
  {journal} {\bibinfo  {journal} {Journal of Computational Physics}\ }\textbf
  {\bibinfo {volume} {117}},\ \bibinfo {pages} {1} (\bibinfo {year}
  {1995})}\BibitemShut {NoStop}%
\bibitem [{noa()}]{noauthor_lammps_nodate}%
  \BibitemOpen
  \href {http://lammps.sandia.gov} {\bibinfo {title} {{LAMMPS} {Official}
  {Website}: http://lammps.sandia.gov}}\BibitemShut {NoStop}%
\bibitem [{\citenamefont {Uhlenbeck}\ and\ \citenamefont
  {Ornstein}(1930)}]{uhlenbeck_theory_1930}%
  \BibitemOpen
  \bibfield  {author} {\bibinfo {author} {\bibfnamefont {G.~E.}\ \bibnamefont
  {Uhlenbeck}}\ and\ \bibinfo {author} {\bibfnamefont {L.~S.}\ \bibnamefont
  {Ornstein}},\ }\href {https://doi.org/10.1103/PhysRev.36.823} {\bibfield
  {journal} {\bibinfo  {journal} {Physical Review}\ }\textbf {\bibinfo {volume}
  {36}},\ \bibinfo {pages} {823} (\bibinfo {year} {1930})}\BibitemShut
  {NoStop}%
\bibitem [{\citenamefont {Green}(1952)}]{green_markoff_1952}%
  \BibitemOpen
  \bibfield  {author} {\bibinfo {author} {\bibfnamefont {M.~S.}\ \bibnamefont
  {Green}},\ }\href {https://doi.org/10.1063/1.1700722} {\bibfield  {journal}
  {\bibinfo  {journal} {The Journal of Chemical Physics}\ }\textbf {\bibinfo
  {volume} {20}},\ \bibinfo {pages} {1281} (\bibinfo {year}
  {1952})}\BibitemShut {NoStop}%
\bibitem [{\citenamefont {Green}(1954)}]{green_markoff_1954}%
  \BibitemOpen
  \bibfield  {author} {\bibinfo {author} {\bibfnamefont {M.~S.}\ \bibnamefont
  {Green}},\ }\href {https://doi.org/10.1063/1.1740082} {\bibfield  {journal}
  {\bibinfo  {journal} {The Journal of Chemical Physics}\ }\textbf {\bibinfo
  {volume} {22}},\ \bibinfo {pages} {398} (\bibinfo {year} {1954})}\BibitemShut
  {NoStop}%
\bibitem [{\citenamefont
  {Kubo}(1957)}]{kubo_ryogo_statistical-mechanical_1957}%
  \BibitemOpen
  \bibfield  {author} {\bibinfo {author} {\bibfnamefont {R.}~\bibnamefont
  {Kubo}},\ }\href@noop {} {\bibfield  {journal} {\bibinfo  {journal} {Journal
  of the Physical Society of Japan}\ }\textbf {\bibinfo {volume} {12}},\
  \bibinfo {pages} {570} (\bibinfo {year} {1957})}\BibitemShut {NoStop}%
\bibitem [{\citenamefont {Weitz}\ \emph {et~al.}(1989)\citenamefont {Weitz},
  \citenamefont {Pine}, \citenamefont {Pusey},\ and\ \citenamefont
  {Tough}}]{weitz_nondiffusive_1989}%
  \BibitemOpen
  \bibfield  {author} {\bibinfo {author} {\bibfnamefont {D.~A.}\ \bibnamefont
  {Weitz}}, \bibinfo {author} {\bibfnamefont {D.~J.}\ \bibnamefont {Pine}},
  \bibinfo {author} {\bibfnamefont {P.~N.}\ \bibnamefont {Pusey}},\ and\
  \bibinfo {author} {\bibfnamefont {R.~J.~A.}\ \bibnamefont {Tough}},\ }\href
  {https://doi.org/10.1103/PhysRevLett.63.1747} {\bibfield  {journal} {\bibinfo
   {journal} {Physical Review Letters}\ }\textbf {\bibinfo {volume} {63}},\
  \bibinfo {pages} {1747} (\bibinfo {year} {1989})}\BibitemShut {NoStop}%
\bibitem [{\citenamefont {Kullback}\ and\ \citenamefont
  {Leibler}(1951)}]{kullback1951information}%
  \BibitemOpen
  \bibfield  {author} {\bibinfo {author} {\bibfnamefont {S.}~\bibnamefont
  {Kullback}}\ and\ \bibinfo {author} {\bibfnamefont {R.~A.}\ \bibnamefont
  {Leibler}},\ }\href@noop {} {\bibfield  {journal} {\bibinfo  {journal} {The
  annals of mathematical statistics}\ }\textbf {\bibinfo {volume} {22}},\
  \bibinfo {pages} {79} (\bibinfo {year} {1951})}\BibitemShut {NoStop}%
\end{thebibliography}%
\newpage

\clearpage

\appendix
\section{Ballistic-Diffusive Distribution Function}
\label{app:ball_diff}
The BDM probability distribution function is defined in Eq.\,(\ref{eq:p_bdm}). We derive the standard deviation $\sigma_\mathrm{diff}$ from the second order Gaussian integral
\begin{widetext}
\begin{equation}
    \begin{split}
        \mu_2 &= \int_{-\infty}^{\infty} P^\mathrm{BDM}(\delta x)(\delta x)^2\,d\delta x\\ 
        &= \int_{-\infty}^{\infty} \exp\left(-\frac{\Delta t}{\tau}\right)P^\mathrm{ball}(\delta x)(\delta x)^2\,d\delta x+\int_{-\infty}^{\infty} \left[1-\exp\left(-\frac{\Delta t}{\tau}\right)\right]P^\mathrm{diff}(\delta x)(\delta x)^2\,d\delta x\\ 
        &= \int_{-\infty}^{\infty} \ddfrac{\exp\left(-\frac{\Delta t}{\tau}\right)}{[2\pi k_BT(\Delta t)^2]^{d/2}}\exp\left(-\frac{(\delta x)^2}{2 k_BT(\Delta t)^2}\right)(\delta x)^2\,d\delta x+\int_{-\infty}^{\infty}\frac{\left[1-\exp\left(-\frac{\Delta t}{\tau}\right)\right]}{[2\pi \sigma_\mathrm{diff}^2]^{d/2}}\exp\left(-\ddfrac{(\delta x)^2}{2 \sigma_\mathrm{diff}^2}\right)(\delta x)^2\,d\delta x\\
        &= \exp\left(-\frac{\Delta t}{\tau}\right)k_BT(\Delta t)^2 + \left[1-\exp\left(-\frac{\Delta t}{\tau}\right)\right]\sigma_\mathrm{diff}^2
    \end{split}
    \label{app:mu_2}
\end{equation}
\end{widetext}
for one dimension ($d=1$). Now, we express the standard deviation of $P^\mathrm{diff}(\delta x)$ as 
\begin{equation}
    \begin{split}
        \sigma_\mathrm{diff} &= \sqrt{\frac{\mu_2 - \exp\left(-\frac{\Delta t}{\tau}\right)k_BT(\Delta t)^2}{\left[1-\exp\left(-\ddfrac{\Delta t}{\tau}\right)\right]}}
    \end{split}
    \label{app_eq:sigma_diff}
\end{equation}
This completes the definition of $P^\mathrm{BDM-T}(X_i)$ using $\mu_2$ and $\sigma_\mathrm{diff}$ recovered by Eqs.\,(\ref{eq:msd}) and (\ref{eq:vel_corr_exp_decay}), respectively. 

In the second version of the BDM model, we match the second and the fourth order moments measured directly from the MD simulations to the probability distribution function. The mean free time $\tau$ is not anymore a function of the velocity auto-correlation function but a free parameter. The derivation of $\sigma_\mathrm{diff}$ in Eqs.\,(\ref{app:mu_2}) and (\ref{app_eq:sigma_diff}) is still valid. In addition, we fit the fourth order moment $\mu_4$ using the fourth order Gaussian integral
\begin{widetext}
\begin{equation}
    \begin{split}
        \mu_4 &= \int_{-\infty}^{\infty} P^\mathrm{BDM}(\delta x)(\delta x)^4\,d\delta x\\ 
        &= \int_{-\infty}^{\infty} \exp\left(-\frac{\Delta t}{\tau}\right)P^\mathrm{ball}(\delta x)(\delta x)^4\,d\delta x+\int_{-\infty}^{\infty} \left[1-\exp\left(-\frac{\Delta t}{\tau}\right)\right]P^\mathrm{diff}(\delta x)(\delta x)^4\,d\delta x\\ 
        &= \frac{3\sqrt{\pi}}{4}\left[\ddfrac{\left(2k_BT(\Delta t)^2\right)^{5/2}}{\frac{\sqrt{2\pi k_BT(\Delta t)^2}}{\exp\left(-\frac{\Delta t}{\tau}\right)}} + \ddfrac{\left(2\sigma_\mathrm{diff}^2\right)^{5/2}}{\frac{\sqrt{2\pi\sigma_\mathrm{diff}^2}}{1-\exp\left(-\frac{\Delta t}{\tau}\right)}}\right]\\ 
        &= \frac{3\exp\left(-\frac{\Delta t}{\tau}\right)(2k_BT(\Delta t)^2)^2\sqrt{2\pi k_BT(\Delta t)^2}}{4\sqrt{2\pi k_BT(\Delta t)^2}}+ \frac{3\left[1-\exp\left(-\frac{\Delta t}{\tau}\right)\right](2\sigma_\mathrm{diff}^2)^2\sqrt{2\pi \sigma_\mathrm{diff}^2}}{4\sqrt{2\pi \sigma_\mathrm{diff}^2}}\\ 
        &= 3\left[\sigma_\mathrm{diff}^4 + \exp\left(-\frac{\Delta t}{\tau}\right)\left((k_BT(\Delta t)^2)^2 - \sigma_\mathrm{diff}^4\right)\right].
\end{split}
\end{equation}
\end{widetext}
Now, we derive the mean free time $\tau$ as a function of the time step $\Delta t$, and the second and the fourth order moments measured from the MD simulation
\begin{equation}
\begin{split}
        \exp\left(-\frac{\Delta t}{\tau}\right) &= \frac{\frac{\mu_4}{3} - \sigma_\mathrm{diff}^4}{(k_BT(\Delta t)^2)^2 - \sigma_\mathrm{diff}^4}\\
        \tau &= \ddfrac{-\Delta t}{\ln{\left(\frac{\frac{\mu_4}{3} - \sigma_\mathrm{diff}^4}{(k_BT(\Delta t)^2)^2 - \sigma_\mathrm{diff}^4}\right)}}.
    \end{split}
    \label{app_eq:tau_sigma_diff}
\end{equation}
Eqs.\,(\ref{app_eq:sigma_diff}) and (\ref{app_eq:tau_sigma_diff}) define a system of linear equations with two unknowns. After substituting Eq.\,(\ref{app_eq:sigma_diff}) in Eq.\,(\ref{app_eq:tau_sigma_diff}), we found a unique solution for $\tau$ given by 
\begin{equation}
\begin{split}
    \exp&\left(-\frac{\Delta t}{\tau}\right) = \ddfrac{\mu_2^2-\frac{\mu_4}{3}}{k_BT(\Delta t)^2\left[2\mu_2-k_BT(\Delta t)^2\right]-\frac{\mu_4}{3}}\\
    \tau &= \ddfrac{-\Delta t}{\ln{\left(\ddfrac{\mu_2^2-\frac{\mu_4}{3}}{k_BT(\Delta t)^2\left[2\mu_2-k_BT(\Delta t)^2\right]-\frac{\mu_4}{3}}\right)}}
\end{split}
\end{equation}
The mean free time $\tau$ is a function of $\mu_2$, $\mu_4$, $\Delta t$ and the temperature of the gas given in LJ units. The standard deviation $\sigma_\mathrm{diff}$ is recovered using Eq.\,(\ref{app_eq:sigma_diff}).

\section{Poisson Weighted Sum of Gaussian Distribution Functions}
\label{app:poisson}
Without collisions particles will move with a constant velocity drawn from a Gaussian distribution function. In this case, the distribution of displacements is given by $P^\mathrm{ball}(X_i)$ in Eq.\,(\ref{eq:p_ball}). If we ought to calculate the distribution of particle displacements for particles that undergo a single collision at a random time $0<t_c<\Delta t$, we would define a sum of two Gaussian distributed random numbers with a second moment given by
\begin{equation}
    t_c^2k_BT + (\Delta t +t_c)^2k_BT=(\Delta t^2 + 2t_c^2-2t_c\Delta t)k_BT
\end{equation}
which is less than the collisionless case except for $t_c=0$ and $t_c=\Delta t$. The full distribution function in one dimension ($d=1$) is then 
\begin{widetext}
\begin{equation}
    \begin{split}
    &P_{\delta x_c}(\delta x)= \\&=\int_{-\infty}^{\infty}\frac{1}{[2\pi k_BTt_c^2]^{d/2}}\exp\left(-\frac{(\delta x_c)^2}{2k_BTt_c^2}\right)\frac{1}{[2\pi k_BT(\Delta t^2 - 2\Delta t\;t_c+2 t_c^2)]^{d/2}}\exp\left(-\frac{(\delta x-\delta x_c)^2}{2k_BT(\Delta t^2 - 2\Delta t\;t_c+2 t_c^2)}\right)d\delta\delta x_c\\
    &=\frac{1}{\sqrt{2\pi k_BT(\Delta t^2 - 2\Delta t\;t_c+2 t_c^2)}}\exp{\left(-\frac{(\delta x)^2}{2k_BT(\Delta t^2 - 2\Delta t\;t_c+2 t_c^2)}\right)}
    \end{split}
    \label{app_eq:P_tc}
\end{equation}
\end{widetext}
where $\delta x_c$ is the displacement for time $0$ to $t_c$, and $(\delta x-\delta x_c)$ for time $t_c$ to $\Delta t$. This results to a Gaussian distribution function with total displacement $\delta x$ for a collision taking place at time $t_c$. To ensure that the time $t_c$ is arbitrary and collisions at any time will be uniformly likely, we average over all possible collision times given by
\begin{widetext}
\begin{equation}
    \begin{split}
        P_{\delta t_c}(\delta x)&=\frac{1}{\Delta t}\int_0^{\Delta t}P_{\delta x_c}(\delta x)\,dt_c\\
        &=\frac{1}{\Delta t}\int_0^{\Delta t}\frac{1}{\sqrt{2\pi k_BT(\Delta t^2 - 2\Delta t\;t_c+2 t_c^2)}}\exp{\left(-\frac{(\delta x)^2}{2k_BT(\Delta t^2 - 2\Delta t\;t_c+2 t_c^2)}\right)}\,dt_c
    \end{split}
    \label{app_eq:random_tc}
\end{equation}
\end{widetext}
It is difficult to evaluate this integral analytically, but it can be solved numerically. However, this is the theory for only one collision occurring at a random time $t_c$. For the Poisson weighted sum of Gaussians in Section\,\ref{sec:poisson_gaussians}, we consider multiple collisions at multiple arbitrary times, which leads to high-dimensional integrals, whose solution is out of the scope of this publication. In addition to the numerical difficulty that multidimensional integrals pose, the resulting probability distribution functions are non-Gaussian. To avoid this, we assume that the collisions are evenly distributed which may introduce a small error.

The WSG probability distribution function is defined in Eq.\,(\ref{eq:poisson_v2}) and recovers the second order moment given by
\begin{widetext}
\begin{equation}
    \begin{split}
        \mu_2 =& \int_{-\infty}^\infty P^\mathrm{WSG}(\delta x)(\delta x)^2 \,d\delta x\\ 
        &= \int_{-\infty}^\infty \sum_{c=0}^\infty e^{-\lambda} \frac{\lambda^c}{c!} \frac{\sqrt{\lambda+1}}{\sqrt{2\pi (c+1)\langle(\delta x)^2\rangle}}\exp\left(-\frac{(\lambda+1)(\delta x)^2}{2(c+1)\langle (\delta x)^2\rangle}\right) (\delta x)^2\,d\delta x\\ 
        &= \sum_{c=0}^\infty e^{-\lambda}\frac{\lambda^c}{c!} \frac{(c+1)\langle(\delta x)^2\rangle}{\lambda+1}\\ 
        &=\langle (\delta x)^2\rangle \frac{e^{-\lambda}}{\lambda+1} \left(\sum_{c=0}^{\infty} \frac{c\lambda^c}{c!}+\sum_{c=0}^{\infty} \frac{\lambda^c}{c!}\right)\\ 
        &=\langle (\delta x)^2\rangle e^{-\lambda} \frac{\lambda}{\lambda+1}\left(\sum_{c=0}^{\infty} \frac{\lambda^c}{c!}+\frac{e^\lambda}{\lambda}\right)\\ 
        &=\langle (\delta x)^2\rangle.
    \end{split}
\end{equation}
Analogously, one derives the fourth order moment as 
\begin{equation}
    \begin{split}
    \mu_4 &= \int_{-\infty}^\infty P(\delta x)(\delta x)^4 \,d(\delta x)\\ 
    &= \int_{-\infty}^\infty \sum_{c=0}^\infty e^{-\lambda} \frac{\lambda^c}{c!} \frac{\sqrt{\lambda+1}}{\sqrt{2\pi (c+1)\langle(\delta x)^2\rangle}}\exp\left(-\frac{(\lambda+1)(\delta x)^2}{2(c+1)\langle(\delta x)^2\rangle}\right) (\delta x)^4 \,d\delta x\\ 
    &= \sum_{c=0}^\infty e^{-\lambda} \frac{\lambda^c}{c!}\left[\ddfrac{3\sqrt{\pi}\left(\frac{2(c+1)\langle(\delta x)^2\rangle}{(\lambda+1)}\right)^{5/2}}{4\sqrt{\frac{2\pi(c+1)\langle(\delta x)^2\rangle}{(\lambda+1)}}}\right]\\ 
    &= \frac{3 \langle(\delta x)^2\rangle^2}{(\lambda+1)^2}\left[e^{-\lambda}\sum_{c=0}^\infty \frac{\lambda^c(c^2+2c+1)}{c!}\right]\\ 
    &=\frac{3\langle(\delta x)^2\rangle^2}{(\lambda+1)^2}\left[e^{-\lambda}\sum_{c=0}^\infty \frac{\lambda^cc^2}{c!}+2e^{-\lambda}\sum_{c=0}^\infty \frac{\lambda^cc}{c!}+e^{-\lambda}\sum_{c=0}^\infty \frac{\lambda^c}{c!}\right]\\ 
    &=\frac{3 \langle(\delta x)^2\rangle^2}{(\lambda+1)^2}\left[e^{-\lambda}\sum_{c=1}^\infty \frac{\lambda^cc}{(c-1)!}+2\lambda e^{-\lambda}\sum_{c=1}^\infty \frac{\lambda^{c-1}}{(c-1)!}+1\right]\\ 
    &=\frac{3 \langle(\delta x)^2\rangle^2}{(\lambda+1)^2}\left[e^{-\lambda}\lambda\frac{d}{d\lambda}\sum_{c=1}^\infty \frac{\lambda^c}{(c-1)!}+2\lambda+1\right]\\ 
    &=\frac{3 \langle(\delta x)^2\rangle^2}{(\lambda+1)^2}\left[e^{-\lambda}\lambda\frac{d}{d\lambda}\lambda e^\lambda+\lambda+1\right]\\ 
    &=\frac{3 \langle(\delta x)^2\rangle^2}{(\lambda+1)^2}\left[e^{-\lambda}\lambda(e^\lambda+\lambda e^\lambda)+2\lambda+1\right]\\
    &=\frac{3 \langle(\delta x)^2\rangle^2}{(\lambda+1)^2}\left[\lambda^2+3\lambda+1\right]
   \end{split}
   \label{app_eq:fourth_order}
\end{equation}
\end{widetext}
This description allows us to adjust $\lambda$, such that the fourth order moment does converge to the measured MD value. We express $\lambda$ as a function of $\Delta t$ with mean squared displacement and fourth order moment measured directly from the MD simulation
\begin{equation}
    \frac{3 \langle(\delta x)^2\rangle^2}{(\lambda+1)^2}\left[\lambda^2+3\lambda+1\right] -\mu_4 = 0
\end{equation}
After solving this quadratic equation, we obtain the following solutions
\begin{equation}
    \lambda_{1,2} = \frac{-9\mu_2^2\pm\sqrt{3[15\mu_2^4-4\mu_2^2\mu_4]}+2\mu_4}{2[3\mu_2^2-\mu_4]}.
\end{equation}
\end{document}